\documentclass{aa}

\usepackage{graphicx}
\usepackage[caption=false]{subfig}
\usepackage{txfonts}
\usepackage[breaklinks=true]{hyperref}
\usepackage{tabularx}
\usepackage{xltabular}
\usepackage{booktabs}
\usepackage{color}
\usepackage{lipsum}
\usepackage[T1]{fontenc}
\usepackage{textcomp, gensymb} 
\usepackage{mathtools}

\usepackage[frozencache=true,cachedir=minted-cache]{minted} 
\usepackage{xcolor} 
\definecolor{LightGray}{gray}{0.9}
\setminted[python]{
baselinestretch=1.2,
bgcolor=LightGray,
fontsize=\footnotesize,
linenos
}

\makeatletter
\renewcommand*\aa@pageof{, page \thepage{} of \pageref*{LastPage}}
\makeatother

\begin{document}

   \title{Adaptive Optics Telemetry Standard}

   \subtitle{Design and specification of a novel data exchange format}

    \author{
            Tiago Gomes\inst{1,2}
        \and
            Carlos M. Correia\inst{1,2}
        \and
            Lisa Bardou\inst{3}
        \and
            Sylvain Cetre\inst{3}
        \and
            Johann Kolb\inst{4}
        \and
            Caroline Kulcsár\inst{5}
        \and
            François Leroux\inst{6}
        \and
            Timothy Morris\inst{3}
        \and
            Nuno Morujão\inst{1,2}
        \and
            Benoît Neichel\inst{6}
        \and
            Jean-Luc Beuzit\inst{6}
        \and
            Paulo Garcia\inst{1,2}
    }
    
    \institute{
            Faculdade de Engenharia da Universidade do Porto, Rua Dr. Roberto Frias, s/n, 4200-465 Porto, Portugal
        \and
            Center for Astrophysics and Gravitation, Instituto Superior Técnico, Av. Rovisco Pais 1, 1049-001 Lisboa, Portugal
        \and
            Durham University (United Kingdom)
        \and
            European Southern Observatory
        \and
            Université Paris-Saclay, Institut d'Optique Graduate School, CNRS, Laboratoire Charles Fabry, 91127 Palaiseau, France
        \and
            Aix Marseille Univ, CNRS, CNES, LAM, Marseille, France
    }

   \date{Received TBD; accepted TBD}

 
  \abstract
   {The amount of Adaptive Optics (AO) telemetry generated by VIS/NIR ground-based observatories is ever greater, leading to a growing need for a standardised data exchange format to support performance analysis and AO research and development activities that involve large-scale telemetry mining, processing, and curation.}
   {This paper introduces the Adaptive Optics Telemetry (AOT) data exchange format as a standard for sharing AO telemetry from visible/infrared ground-based observatories. AOT is based on the Flexible Image Transport System (FITS) and aims to provide unambiguous and consistent data access across various systems and configurations, including natural and single/multiple laser guide-star AO systems.}
   {We designed AOT focused on two key use cases: atmospheric turbulence parameter estimation and point-spread function reconstruction (PSF-R). We prototyped and tested the design using existing AO telemetry datasets from multiple systems: single conjugate with natural and laser guide stars, tomographic systems with multi-channel wavefront sensors, single and multi wavefront correctors in systems featuring either a Shack-Hartmann or Pyramid as main wavefront sensors.}
   {The AOT file structure has been thoroughly defined, specifying data fields, descriptions, data types, units, and expected dimensions. To support this format, we have developed a Python package that enables data conversion, reading, writing and exploration of AOT files, which has been made publicly available and compatible with a general-purpose Python package manager. We demonstrate the flexibility of the AOT format by packaging data from five different instruments, installed on different telescopes.}
   {}
    \keywords{Adaptive Optics --
            Telemetry --
            Data Exchange Format --
            FITS --
            Data Model --
            PSF reconstruction --
            Atmospheric turbulence 
    }

   \maketitle
%

\section{Introduction} \label{s:introduction}
The adoption and interest in Adaptive Optics (AO) for astronomy has increased significantly in recent years. As a consequence, there has been a large increase in observatories and researchers producing AO telemetry data. We use the term ``telemetry'' to represent AO internal signals such as wavefront sensor measurements, deformable mirror commands and several other reconstruction and control matrices and parameters.

Recent developments have shown that AO telemetry data is of high scientific interest, allowing for use-cases such as:
\begin{itemize}
    \item Advanced exploitation of scientific data - telemetry data can be used to derive the state of the atmospheric turbulence during observations \citep{Vidal-a-10, guesalaga2014using, Martin2016WHT, Jolissaint:18, Laidlaw2019Automated, andrade2019estimation}. For situations where the point-spread function (PSF) of the system is not available it allows to derive one, with high accuracy \citep{1997JOSAA..14.3057V, 2006A&A...457..359G, gilles2012simulation, 2019MNRAS.487.5450B, 2020A&A...635A.208F, 2020MNRAS.494..775B}. These situations are very common nowadays, and related to high impact science cases.
    \item System performance estimation and runtime optimization - applications such as fine-tuning of AO systems, advanced real-time strategies, wave-front prediction or even scheduling of observations \citep{sivo2014first, petit2014sphere, sinquin2020sky}.
    \item Instrumentation research - for either established or newer teams without direct access to AO instrumentation \citep{conan2014object, 10.1117/12.2597170}.
\end{itemize}

Despite the demonstrated usefulness as a science product, AO telemetry has been traditionally seen as ``engineering data'', typically used internally for one-off instrument commissioning or regular calibration \citep{hirst2020telemetry}. As a result of this, AO telemetry datasets are usually saved in private archives that are out of reach for the end-user, or they may simply not be kept in the long-term. Even in cases where third-parties can access telemetry datasets, their documentation tends to be poor or non-existent. This is problematic because each data-producing system may be saving different sets of data, using different formats and providing access through different ways (in some cases, even datasets generated by the same system are completely distinct). As a result, the task of interpreting and analysing data is a huge burden on the users and it is difficult to create data analysis programs that can be generalised to multiple systems.

In light of this context, we consider that AO telemetry currently has a large untapped potential that can be uncovered by broadening the access to this data, in what we call the ``hidden iceberg'' of adaptive optics. Accordingly, many systems are starting to show interest in saving and sharing such data. One of the earliest examples of public archiving of AO telemetry was carried out by the CANARY project \citep{canary2008}, having made its open loop data for the sodium laser guide star experiment \citep{bardou2021canary} available on the ESO Science Archive, using a data format that is extensively detailed in an accompanying manual \citep{canary_archive, canary_manual}. The Gemini Observatory is planning on publishing one of the first large scale archives of telemetry data \citep{hirst2020telemetry} (mostly composed of wavefront slopes and DM commands), through the Gemini Observatory Archive \citep{hirst2016archive}. However, publishing AO telemetry data currently requires significant effort from the responsible teams, as they are forced to design their own purpose-made data format, and make it available along with proper documentation for their prospective users. While such formats represent a significant step forward in data accessibility, given that they are tailored specifically for one instrument/observatory, they typically make no attempts at generalising support for multiple systems. In fact, to our knowledge, there has been no major attempt in the AO community to establish a consensual data format for AO telemetry to date.

We believe that access to AO telemetry data can only become ubiquitous by having the bulk of the AO community agree on a common way of sharing their data, that is, agreeing on a data exchange format. A data exchange format essentially defines a set of attributes/fields, their meanings, the way they relate to other data and the way they are stored in a shareable format. We share our considerations on designing such a format in Sec.\ref{s:considerations}.

With these considerations in mind, we propose the Adaptive Optics Telemetry (AOT) data exchange format, built on top of the Flexible Image Transport System (FITS) \citep{wells_fits-flexible_1979}. This format draws some inspiration from OIFITS \citep{pauls_data_2004, pauls_data_2005, duvert_oifits_2017}, which was created as a reaction to similar challenges that the interferometry community was facing. This paper aims to discuss the thought process behind the design decisions and to standardise this format by fully defining its details such as its structure, the set of data that may be shared and their respective definitions, conventions and units (Sec.~\ref{ss:data_format}). In Sec.\ref{sec:tools}, we also discuss how tools and data models can support the format by providing useful abstractions for the users, on which a previous paper has been published \citep{towards_SPIE_2022}.

In summary, AOT aims to provide a common interface for AO telemetry data access regardless of the specific data-producing system, its size (VLT-class, ELT-class), the AO mode used (SCAO, SLAO, GLAO, MOAO, LTAO, MCAO), number and type of wavefront sensors (Shack-Hartmann, Pyramid) and specific wavefront correction setup  (linear stages, tip-tilt mirrors, deformable mirrors). It is envisioned as a convenient way of gathering relevant data from a single system (which may have been produced by different instruments/devices, in different subsystems) and packaging everything in a easily shareable format that is well-understood. This means that AOT is not meant to replace the actual real-time writing (dumping) of instrument data, but rather to be used as a data processing step that makes such data digestible by its users. While the format defines a large amount of fields that can describe the different aspects of telemetry data, most of these are not mandatory; thus file sizes vary significantly based on the needs of the user, the complexity of the system and data reduction mechanisms that may have been applied. 

This project is being developed in the context of OPTICON–RadioNet Pilot, an European collaboration through which we aim to promote broad discussion and consensus within the AO community. A set of proof-of-concept AOT files has been made publicly available on both the ESO archive (for existing ESO systems) and Zenodo.

The emphasized words "\textit{must}", "\textit{must not}", "\textit{should}", "\textit{should not}", "\textit{recommended}",  "\textit{may}", and "\textit{optional}" in this document are to be interpreted as described in RFC 2119 \citep{rfc2119}.

\section{Format Considerations} \label{s:considerations}
The overall goal of this data exchange format is to lay the foundation to allow easier data sharing of any interested teams and researchers in the future. In other words, we aim to support the community's need of treating AO telemetry data as a science product. To achieve this, we need to create a format that is widely used by the community. Since this requires support by the many interested parts in the community, we paid special attention in defining the set of use-cases that we want to support (Sec.~\ref{ss:use_cases}) and a data model that would streamline such usage (Sec.~\ref{ss:data_model}). We also pondered on the suitable file formats for implementing that data model (Sec.~\ref{ss:file_format}). Overall, our decision-making was guided by the design principles set in Appendix~\ref{app:design_principles}.

\subsection{Use-cases} \label{ss:use_cases}
To ensure that a format fits the user's needs, it is essential to design it around the specific use-cases it is intended to support, as that allows us to verify its completeness for those purposes. In other words, these use-cases enable a top-down approach to the design, as they directly imply a set of requirements that must be satisfied. Although AO telemetry data has a wide range of use-cases in astronomy, AOT will be designed with two main use-cases in mind, which we intend to fully support: 1) atmospheric turbulence parameters estimation and 2) point-spread function reconstruction (PSF-R). 

We have looked into state-of-the-art algorithms for these use-cases to guarantee that, to our knowledge, AOT is able to package all data that may be relevant. AOT also specifies some data that, although not currently used for these purposes, we expect may be necessary for future algorithms. Finally, AOT can accommodate data that is not directly targeted at these specific use-cases, but that may help broaden the format's usefulness and potential applications, or simply improve its convenience and user-friendliness. Despite our primarily top-down approach, we also made sure to analyse datasets already being created by existing systems; this bottom-up approach allowed us to identify the types of data that the community currently prioritises and ensure AOT can fill those needs.

It is important to note that science images are \textbf{not} a part of AOT. These are already commonly shared within the astronomical community through fairly well-established means, therefore there does not seem to be a benefit in sharing them through a different format. Regardless, public data archives may choose to provide science data along with AOT files, in separate files.

\subsection{Data model} \label{ss:data_model}
\begin{figure*}[htbp]
   \begin{center}
   \includegraphics[width=\textwidth]{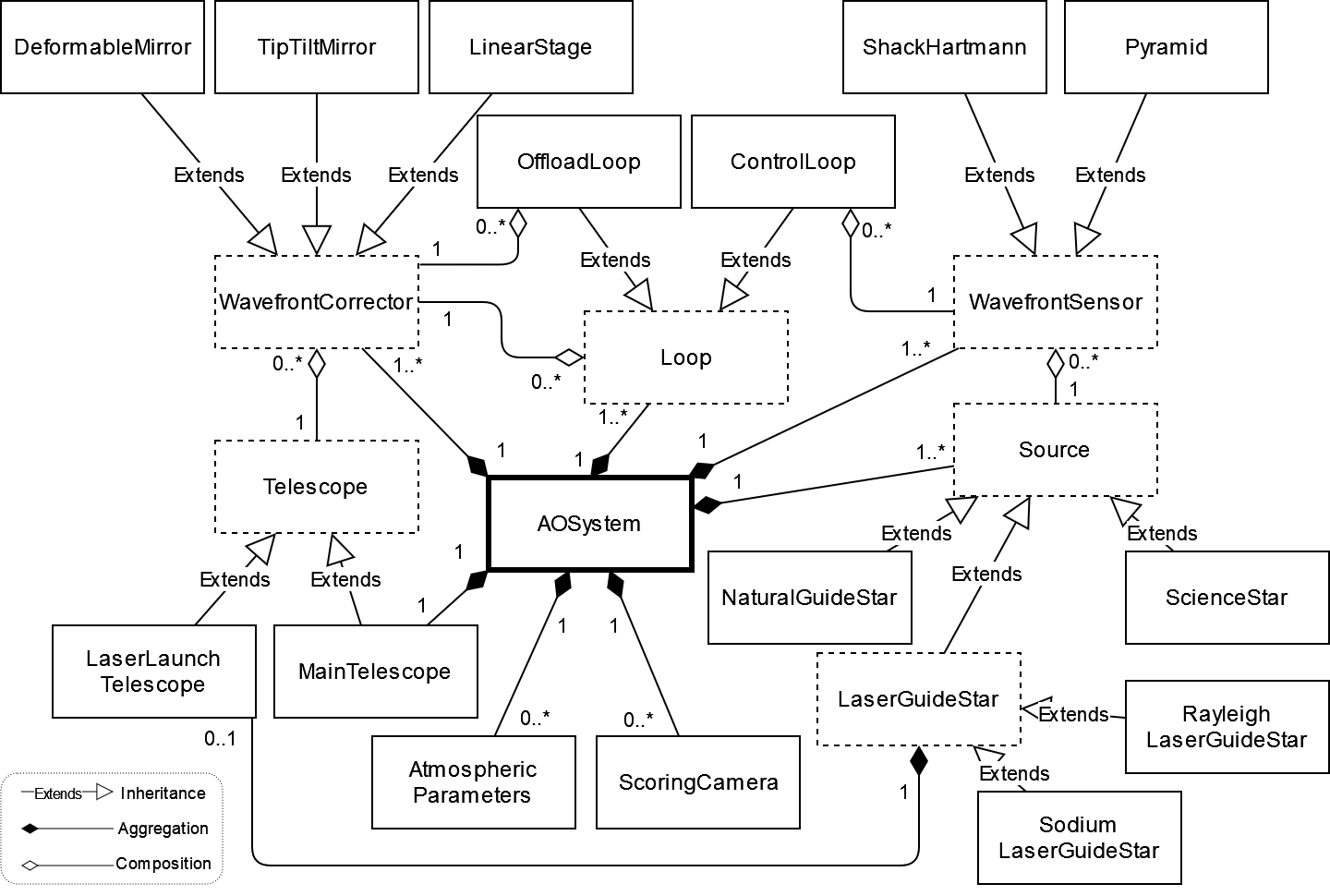}
   \end{center}
   \caption{\label{fig:umlclassdiagram} 
    Simplified UML Class Diagram representing the AOT data model. The central \textit{AOSystem} class can hold information about the entire system, via references to its multiple parts. Classes represented via dashed rectangles are abstract, and are implemented by the classes that inherit from them. For simplicity, some inner classes have been omitted.}
\end{figure*}

In the development of AOT, we designed a data model that attempts to generalise the different parts that may compose an AO system and their relationships, such that it can be used to describe a wide variety of system configurations (including Single Conjugate, Single Laser, Ground Layer, Multi-Object, Laser Tomography and Multi-Conjugate Adaptive Optics) and different wavefront correction and sensing setups. This data model defines conceptual relationships between the different sets of instruments and their data, with the goal of providing an intuitive and simplified way of accessing data. An overview of this data model can be seen in Fig.~\ref{fig:umlclassdiagram}.

The AOT data model is agnostic both in terms of programming language and specific file format, meaning it can be used as the basis for any implementation of the format. An earlier version of this data model has been discussed in greater detail in \cite{towards_SPIE_2022}.

\subsection{File format} \label{ss:file_format}
While the AOT data model does not imply any specific file format, in this paper we will provide the reference implementation of this data model in a specific file format. Rather than creating our own file format, we believe it is in the best interest of the astronomical community to opt for one that is already commonly used in the community, as that allows us to leverage the high degree of familiarity that astronomers already have with such file formats and the widely available tools for creating, reading and editing them. Therefore, we looked into established file formats that would be able to contain all the necessary data while being easily shareable, namely: FITS \citep{wells_fits-flexible_1979}, HDF5 \citep{hdf5}, VOTable \citep{ochsenbein_votable_2004} and ASDF \citep{greenfield_asdf_2015}.

Ultimately, we opted to provide an implementation using the FITS standard. The biggest driving factor in this decision was that we aimed to publish AOT data on the ESO Science Archive, which currently only provides full support for this format. Nonetheless, we acknowledge that FITS is an ageing standard that can be a limiting in some particular scenarios \citep{thomas_learning_2015, scroggins_once_2020}; therefore there might be interest in implementing the AOT data model using other file formats in the future.

Significant familiarity with the FITS specification is necessary to fully understand the AOT specification described in the following sections; with that in mind, we provide an overview of the most relevant FITS concepts in Appendix~\ref{app:fits_overview}.

\section{Data Format} \label{ss:data_format}
A standard AOT file fully adheres to the FITS specification; in other words, a file that follows the AOT standard is also a standard FITS file. An AOT file is composed of one specially formatted primary HDU with no data array (Sec.~\ref{ss:primary}), followed by a specific set of binary table extensions (Sec.~\ref{ss:bintables}) and possibly some image extensions (Sec.~\ref{ss:images}). A single AOT file is capable of supporting all relevant information related to a single telemetry recording. In the following sections, we use a set of AOT-specific conventions (see Appendix~\ref{app:conventions}) that help describing the data format in detail.

\subsection{Primary HDU} \label{ss:primary}
The primary HDU in an AOT file contains no data array. However, its header contains a set of keywords that are used to provide metadata about the file and its data in general, as described in Table~\ref{tab:PrimaryHeader}. It is \textit{recommended} that these keywords are defined in the sequence described in the table, after all the mandatory FITS keyword. Keywords that are not strictly mandatory \textit{may} be omitted or have null values.

\begin{table}[htbp] \footnotesize
\caption{\label{tab:PrimaryHeader}Primary Header of an AOT file.}
\centering
\begin{tabularx}{\linewidth}{>{\ttfamily}llX}
\toprule
\normalfont{\textbf{Keyword}} & \textbf{Type} & \textbf{Description}\\
\midrule
AOT-VERS    &\textit{str}   &Its presence identifies the file as an AOT file and indicates the version of the format that it complies with, following Semantic Versioning \citep{prestonwerner2013semantic}. For the current specification, the value \textit{must} be \texttt{'2.0.0'}.\tablefootmark{a}\\
AO-MODE     &\textit{str}   &Describes the system's AO configuration. \textit{Must} be one of \texttt{'SCAO'}, \texttt{'SLAO'}, \texttt{'GLAO'}, \texttt{'MOAO'}, \texttt{'LTAO'} or \texttt{'MCAO'} (which respectively stand for Single Conjugate, Single Laser, Ground Layer, Multi-Object, Laser Tomography and Multi-Conjugate Adaptive Optics).\tablefootmark{a}\\
TIMESYS     &\textit{str}   &Defines the temporal reference frame for the entire file. Value \textit{must} be \texttt{'UTC'}.\tablefootmark{a,b}\\
DATE-BEG    &\textit{str}   &Start time of data acquisition in ISO derived format (see Appendix~\ref{sapp:time}).\tablefootmark{b}\\
DATE-END    &\textit{str}   &Stop time of data acquisition in ISO derived format (see Appendix~\ref{sapp:time}).\tablefootmark{b}\\
SYS-NAME    &\textit{str}   &Name of the AO system that produced the data.\\
STREHL-R    &\textit{float} &Estimated Strehl ratio (arcsec).\\
FITT-ERR    &\textit{float} &Estimated fitting error, in squared radians at the sensing wavelength.\\
ALIA-ERR    &\textit{float} &Estimated aliasing error, in squared radians at the sensing wavelength.\\
NOIS-ERR    &\textit{float} &Estimated photon/detector noise error, in squared radians at the sensing wavelength.\\
TEMP-ERR    &\textit{float} &Estimated temporal error, in squared radians at the sensing wavelength.\\
OTHE-ERR    &\textit{float} &Estimated other sources of error, in squared radians at the sensing wavelength.\\
CONFIG      &\textit{str}   &Free-form text that describes configuration parameters of the system.\\
\bottomrule
\end{tabularx}
\tablefoot{\\
    \tablefoottext{a}{Strictly mandatory (value \textit{must not} be null).}
    \tablefootmark{b}{Standard FITS keyword.}
}
\end{table}

\subsection{Binary tables} \label{ss:bintables}
After the primary HDU, all AOT files files \textit{should} contain a sequence of up to 17 specifically formatted binary table extensions. The names (\texttt{EXTNAME}) of these tables \textit{must} be: 
\begin{enumerate}
    \item \textbf{AOT\_TIME}
    \item \textbf{AOT\_ATMOSPHERIC\_PARAMETERS}
    \item \textbf{AOT\_ABERRATIONS}
    \item \textbf{AOT\_TELESCOPES}
    \item \textbf{AOT\_SOURCES}
    \item AOT\_SOURCES\_SODIUM\_LGS
    \item AOT\_SOURCES\_RAYLEIGH\_LGS
    \item \textbf{AOT\_DETECTORS}
    \item \textbf{AOT\_SCORING\_CAMERAS}
    \item \textbf{AOT\_WAVEFRONT\_SENSORS}
    \item AOT\_WAVEFRONT\_SENSORS\_SHACK\_HARTMANN
    \item AOT\_WAVEFRONT\_SENSORS\_PYRAMID
    \item \textbf{AOT\_WAVEFRONT\_CORRECTORS}
    \item AOT\_WAVEFRONT\_CORRECTORS\_DM
    \item \textbf{AOT\_LOOPS}
    \item AOT\_LOOPS\_CONTROL
    \item AOT\_LOOPS\_OFFLOAD
\end{enumerate}

The tables in bold are the ''mandatory tables'' and they \textit{must} always exist in the file, even if they contain no rows. The remaining are ''secondary tables'' and they are related to the mandatory tables with which they share part of the name (e.g. AOT\_LOOPS\_CONTROL is related to AOT\_LOOPS). If an entry in an secondary table shares the \texttt{UID} with an entry in the mandatory table it is related to, then both entries describe the same object: in that case, the mandatory table holds general data about the object, while the secondary table stores data corresponding to the object's specific sub-type. All secondary tables are \textit{optional}, meaning that they \textit{may} be omitted if they would otherwise be empty (that is, if no elements of that particular sub-type exist).

It is strongly \textit{recommended} that the binary tables in AOT files follow the specified sequence; they are ordered in this way in order to ensure that, when reading sequentially, tables that \textit{may} be referenced by entries on a certain field have always been read previously. However, the order of the rows inside each table is irrelevant; tables that have the same rows ordered differently are considered equivalent.

In the following sections, we will define the purpose of each of these binary table extensions and their contents. It is \textit{recommended} that all of the specified table fields exist (even if all of their entries are null) and that they follow their defined sequence. An AOT file \textit{must} adhere to the name and type of each field defined in the tables of these sections (through the \texttt{TTYPE}\(n\), \texttt{TFORM}\(n\) keywords, respectively). It is \textit{recommended} that the data follows the unit of measurement specified in the table and that it is specified through the \texttt{TUNIT}\(n\) keyword; if an alternative unit of measurement is used, it must be specified through that same keyword. The description of each field (provided in the tables) \textit{may} be saved in the AOT file through a \texttt{TCOMM}\(n\) keyword.

\subsubsection{AOT\_TIME}
The fields of this binary table are described in Table~\ref{tab:AOTTime}.

This table allows the user to relate time-varying data to their corresponding progress of time. Specifically, each row in this table specifies moments at which some data-points occurred, allowing fields or image extension that describe time-varying data to then directly reference their corresponding row.

The time instants in this table \textit{must} be defined through synchronised Unix timestamps and/or through a synchronised frame number, such that two data points that occurred simultaneously always share the same timestamp/frame number. If in one row both a list of timestamps and frame numbers are provided, they \textit{must} share the same length.

\begin{table}[htbp] \footnotesize
\caption{\label{tab:AOTTime}AOT\_TIME fields.}
\centering
\begin{tabularx}{\linewidth}{>{\ttfamily}lllX}
\toprule
\normalfont{\textbf{Name}} & \textbf{Type} & \textbf{Unit} & \textbf{Description}\\
\midrule
UID                 &\textit{str}   &\textit{n.a.}  &Unique ID that identifies the set of time data.\tablefootmark{a,b}\\
TIMESTAMPS          &\textit{lst}   &s              &List of Unix timestamps at which the respective data applies (see Appendix~\ref{sapp:time}).\tablefootmark{c}\\
FRAME\_NUMBERS      &\textit{lst}   &count          &List of frame numbers at which the respective data applies.\tablefootmark{c}\\
\bottomrule
\end{tabularx}
\tablefoot{\\
    \tablefoottext{a}{Strictly mandatory (value \textit{must not} be null).}
    \tablefoottext{b}{All entries in this field \textit{must} be unique.}
    \tablefoottext{c}{For each row in the table, at least one of the entries in these fields \textit{must} not be null; if both are not null, they \textit{must} have the same length.}
}
\end{table}

\subsubsection{AOT\_ATMOSPHERIC\_PARAMETERS}
The fields of this binary table are described in Table~\ref{tab:AOTAtmosphericParameters}.

Each row in this table describes a set of relevant atmospheric parameters that were obtained from a certain data source. Data source is identified via the \texttt{UID} field and the reference wavelength of the data is defined in \texttt{WAVELENGTH}. Then, time data \textit{may} be associated via \texttt{TIME\_UID} and, for each of those instances, data regarding the overall seeing and conditions of each turbulence layer \textit{may} also be defined.

\begin{table}[htbp] \footnotesize
\caption{\label{tab:AOTAtmosphericParameters}AOT\_ATMOSPHERIC\_PARAMETERS fields.}
\centering
\begin{tabularx}{\linewidth}{>{\ttfamily}lllX}
\toprule
\normalfont{\textbf{Name}} & \textbf{Type} & \textbf{Unit} & \textbf{Description}\\
\midrule
UID                     &\textit{str}   &\textit{n.a.}      &Unique ID that identifies the data source for these atmospheric parameters.\tablefootmark{a,b}\\
WAVELENGTH              &\textit{flt}   &m                  &Reference wavelength.\\
TIME\_UID               &\textit{str}   &\textit{n.a.}      &AOT\_TIME row-reference.\\
R0                      &\textit{lst}   &m                  &List of Fried parameters at reference wavelength at zenith, over time.\\
SEEING                  &\textit{lst}   &arcsec             &List of full width at half maximum measures of the seeing disc at reference wavelength at zenith, over time.\\
TAU0                    &\textit{lst}   &s                  &List of atmospheric coherence times at reference wavelength at zenith, over time.\\
THETA0                  &\textit{lst}   &rad                &List of isoplanatic angles at reference wavelength, over time.\\
LAYERS\_REL\_WEIGHT     &\textit{str}   &\textit{n.a.}      &Image-reference.\\
LAYERS\_HEIGHT          &\textit{str}   &\textit{n.a.}      &Image-reference.\\
LAYERS\_L0              &\textit{str}   &\textit{n.a.}      &Image-reference.\\
LAYERS\_WIND\_SPEED     &\textit{str}   &\textit{n.a.}      &Image-reference.\\
LAYERS\_WIND\_DIRECTION &\textit{str}   &\textit{n.a.}      &Image-reference.\\
TRANSFORMATION\_MATRIX  &\textit{str}   &\textit{n.a.}      &Image-reference.\\
\bottomrule
\end{tabularx}
\tablefoot{\\
    \tablefoottext{a}{Strictly mandatory (value \textit{must not} be null).}
    \tablefoottext{b}{All entries in this field \textit{must} be unique.}
}
\end{table}

\subsubsection{AOT\_ABERRATIONS}
The fields of this binary table are described in Table~\ref{tab:AOTAberrations}.

This table defines data that allows for the description of optical aberrations that \textit{may} exist in different parts of the AO system. Rows \textit{may} be referenced by Wavefront Sensors, Scoring Cameras or Wavefront Correctors, and they contain data that defines a set of aberrations related to that device. 

In each row, the \texttt{MODES} field provides a reference to a 3D image, which is interpreted as a set of \(n\) 2D images representing the orthonormal basis of modes for this row. Then, the \texttt{COEFFICIENTS} field provides a reference to a 2D image that contains \(n\) columns, each containing the coefficient for that particular basis, with as many rows as necessary to describe the different aberrations that \textit{may} occur at different field offsets. The field offsets for these aberrations are specified by the \texttt{X\_OFFSETS} and \texttt{Y\_OFFSETS} lists, which provide an horizontal and vertical offset for each set of coefficients. For each field offset, the user is able to calculate the overall optical aberration by multiplying each mode basis by its respective coefficients for that field offset, and summing those results.

If no field offsets are specified, there \textit{may} only be one row of coefficients, which is assumed to be representative of the average field offset.

\begin{table}[htbp] \footnotesize
\caption{\label{tab:AOTAberrations}AOT\_ABERRATIONS fields.}
\centering
\begin{tabularx}{\linewidth}{>{\ttfamily}lllX}
\toprule
\normalfont{\textbf{Name}} & \textbf{Type} & \textbf{Unit} & \textbf{Description}\\
\midrule
UID             &\textit{str}   &\textit{n.a.}  &Unique ID that identifies the aberration.\tablefootmark{a,b}\\
MODES           &\textit{str}   &\textit{n.a.}  &Image-reference.\tablefootmark{a}\\
COEFFICIENTS    &\textit{str}   &\textit{n.a.}  &Image-reference.\tablefootmark{a}\\
X\_OFFSETS      &\textit{lst}   &rad            &List of horizontal offsets from the centre of the field.\tablefootmark{c}\\
Y\_OFFSETS      &\textit{lst}   &rad            &List of vertical offsets from the centre of the field.\tablefootmark{c}\\
\bottomrule
\end{tabularx}
\tablefoot{\\
    \tablefoottext{a}{Strictly mandatory (value \textit{must not} be null).}
    \tablefoottext{b}{All entries in this field \textit{must} be unique.}
    \tablefoottext{c}{For each row in the table, entries in these fields \textit{must} have the same length, equal to the number of sets of coefficients specified.}
}
\end{table}

\subsubsection{AOT\_TELESCOPES}
The fields of this binary table are described in Table~\ref{tab:AOTTelescopes}. 

This table describes all telescopes in the AO system, with each one being either a Main Telescope (that is, the telescope performing the observation itself) or a Laser Launch Telescope (LLT, which generates one or more laser guide stars). There \textit{must} be exactly one Main Telescope, and all LLTs \textit{should} be referenced by at least one laser guide star.

\begin{table*}[htbp] \footnotesize
\caption{\label{tab:AOTTelescopes}AOT\_TELESCOPES fields.}
\centering
\begin{tabularx}{\linewidth}{>{\ttfamily}lllX}
\toprule
\normalfont{\textbf{Name}} & \textbf{Type} & \textbf{Unit} & \textbf{Description}\\
\midrule
UID             &\textit{str}   &\textit{n.a.}  &Unique ID that identifies the telescope.\tablefootmark{a,b}\\
TYPE            &\textit{str}   &\textit{n.a.}  &Indicates the type of telescope (either \texttt{'Main Telescope'} or \texttt{'Laser Launch Telescope'}).\tablefootmark{a}\\
LATITUDE        &\textit{flt}   &deg            &Latitude of the telescope (World Geodetic System).\\
LONGITUDE       &\textit{flt}   &deg            &Longitude of the telescope (World Geodetic System).\\
ELEVATION       &\textit{flt}   &deg            &Elevation of the telescope at start, the angle between the object and the observer's local horizon with 0\degree\ being the horizon and 90\degree\ being zenith (horizontal coordinate system).\\
AZIMUTH         &\textit{flt}   &deg            &Azimuth of the telescope at start, the angle of the object around the horizon with 0\degree\ being North, increasing eastward (horizontal coordinate system).\\
PARALLACTIC     &\textit{flt}   &deg            &Parallactic angle, the spherical angle between the hour circle and the great circle through a celestial object and the zenith.\\
PUPIL\_MASK     &\textit{str}   &\textit{n.a.}  &Image-reference.\\
PUPIL\_ANGLE    &\textit{flt}   &rad            &Clockwise rotation of the pupil mask.\\
ENCLOSING\_D    &\textit{flt}   &m              &Diameter of the smallest circle that contains the entirety of the pupil (enclosing circle).\\
INSCRIBED\_D    &\textit{flt}   &m              &Diameter of the largest circle that can be contained in the pupil (inscribed circle). On monolithic circular pupils this is equivalent to \texttt{ENCLOSING\_D}.\\
OBSTRUCTION\_D  &\textit{flt}   &m              &Diameter of the smallest circle that contains the entire central obstruction.\\
SEGMENTS\_TYPE  &\textit{str}   &\textit{n.a.}  &Describes the type of the segments that constitute the pupil. Value \textit{must} be \texttt{'Monolithic'} if the pupil is monolithic, \texttt{'Hexagon'} if the segments are hexagons or \texttt{'Circle'} if the segments are circles.\tablefootmark{a}\\
SEGMENTS\_SIZE  &\textit{flt}   &m              &Size of the segments that constitute the pupil, measured as the diameter of the smallest circle that contains the entirety of a segment. If no segments exist (monolithic) this \textit{must} be null.\\
SEGMENTS\_X     &\textit{lst}   &m              &List of horizontal coordinates of each segment in the pupil.\tablefootmark{c}\\
SEGMENTS\_Y     &\textit{lst}   &m              &List of vertical coordinates of each segment in the pupil.\tablefootmark{c}\\
TRANSFORMATION\_MATRIX  &\textit{str}   &\textit{n.a.}      &Image-reference.\\
ABERRATION\_UID &\textit{str}   &\textit{n.a.}  &AOT\_ABERRATIONS row-reference.\\
\bottomrule
\end{tabularx}
\tablefoot{\\
    \tablefoottext{a}{Strictly mandatory (value \textit{must not} be null).}
    \tablefoottext{b}{All entries in this field \textit{must} be unique.}
    \tablefoottext{c}{List \textit{must} have the same length as the total number of segments (or be empty if the pupil is monolithic). The origin of the coordinates is the centre of the inscribed circle.}
}
\end{table*}

\subsubsection{AOT\_SOURCES}
The fields of this binary table are described in Table~\ref{tab:AOTSources}.

This table describes all light sources in the system, which \textit{may} be of 4 different types: Science Star, Natural Guide Star (NGS), Sodium Laser Guide Star or Rayleigh Laser Guide Star. Astronomical coordinates for the sources \textit{may} be provided. Both types of Laser Guide Star (LGS) sources \textit{may} provide secondary data through the Tables \ref{tab:AOTSourcesSodiumLGS} or \ref{tab:AOTSourcesRayleighLGS}, respectively.

\begin{table}[htbp] \footnotesize
\caption{\label{tab:AOTSources}AOT\_SOURCES fields.}
\centering
\begin{tabularx}{\linewidth}{>{\ttfamily}lllX}
\toprule
\normalfont{\textbf{Name}} & \textbf{Type} & \textbf{Unit} & \textbf{Description}\\
\midrule
UID                             &\textit{str}   &\textit{n.a.}  &Unique ID that identifies the light source.\tablefootmark{a,b}\\
TYPE                            &\textit{str}   &\textit{n.a.}  &Indicates the type of light source (either \texttt{'Science Star'}, \texttt{'Natural Guide Star'}, \texttt{'Sodium Laser Guide Star'} or \texttt{'Rayleigh Laser Guide Star'}).\tablefootmark{a}\\
RIGHT\_ASCENSION                &\textit{flt}   &deg            &Right ascension of the light source, epoch J2000 (equatorial coordinate system).\\
DECLINATION                     &\textit{flt}   &deg            &Declination of the light source, epoch J2000 (equatorial coordinate system).\\
ELEVATION\_OFFSET               &\textit{flt}   &deg            &Offset from the Main Telescope's \texttt{ELEVATION}.\\
AZIMUTH\_OFFSET                 &\textit{flt}   &deg            &Offset from the Main Telescope's \texttt{AZIMUTH}.\\
WIDTH                           &\textit{flt}   &rad            &Effective width at zenith.\\
\bottomrule
\end{tabularx}
\tablefoot{\\
    \tablefoottext{a}{Strictly mandatory (value \textit{must not} be null).}
    \tablefoottext{b}{All entries in this field \textit{must} be unique.}
}
\end{table}

\begin{table}[htbp] \footnotesize
\caption{\label{tab:AOTSourcesSodiumLGS}AOT\_SOURCES\_SODIUM\_LGS fields.}
\centering
\begin{tabularx}{\linewidth}{>{\ttfamily}lllX}
\toprule
\normalfont{\textbf{Name}} & \textbf{Type} & \textbf{Unit} & \textbf{Description}\\
\midrule
UID             &\textit{str}   &\textit{n.a.}  &Unique ID that identifies the light source.\tablefootmark{a,b}\\
HEIGHT          &\textit{flt}   &m              & Mean LGS height above sea level at zenith.\\
PROFILE         &\textit{str}   &\textit{n.a.}  & Image-reference.\\
ALTITUDES       &\textit{lst}   &m              & LGS layer profile altitudes at zenith. \textit{Must} be the same length as the number of layers defined in \texttt{PROFILE}.\\
LLT\_UID        &\textit{str}   &\textit{n.a.}  &AOT\_TELESCOPES row-reference, which contains data related to the LLT used to generate the source. Referenced row \textit{must} have \texttt{TYPE='Laser Launch Telescope'}.\\
\bottomrule
\end{tabularx}
\tablefoot{\\
    Secondary table exclusively for sources of type \texttt{'Sodium Laser Guide Star'}. If there are entries of this type in the mandatory table, there \textit{must} also exist an entry with the same \texttt{UID} in this table.\\
    \tablefoottext{a}{Strictly mandatory (value \textit{must not} be null).}
    \tablefoottext{b}{All entries in this field \textit{must} be unique.}
}
\end{table}

\begin{table}[htbp] \footnotesize
\caption{\label{tab:AOTSourcesRayleighLGS}AOT\_SOURCES\_RAYLEIGH\_LGS fields.}
\centering
\begin{tabularx}{\linewidth}{>{\ttfamily}lllX}
\toprule
\normalfont{\textbf{Name}} & \textbf{Type} & \textbf{Unit} & \textbf{Description}\\
\midrule
UID         &\textit{str}   &\textit{n.a.}  & Unique ID that identifies the light source.\tablefootmark{a,b}\\
DISTANCE    &\textit{flt}   &m              & Fixed distance of the LGS from the telescope.\\
DEPTH       &\textit{flt}   &m              & Range covered by the laser light while the shutter is opened.\\
LLT\_UID    &\textit{str}   &\textit{n.a.}  & AOT\_TELESCOPES row-reference, which contains data related to the LLT used to generate the source. Referenced row \textit{must} have \texttt{TYPE='Laser Launch Telescope'}.\\
\bottomrule
\end{tabularx}
\tablefoot{\\
    Secondary table exclusively for sources of type \texttt{'Rayleigh Laser Guide Star'}. If there are entries of this type in the mandatory table, there \textit{must} also exist an entry with the same \texttt{UID} in this table.\\
    \tablefoottext{a}{Strictly mandatory (value \textit{must not} be null).}
    \tablefoottext{b}{All entries in this field \textit{must} be unique.}
}
\end{table}

\subsubsection{AOT\_DETECTORS}
The fields of this binary table are described in Table~\ref{tab:AOTDetectors}.

Data related to the physical characteristics and configurations of a detector, as well as the actual pixels recorded by it, \textit{may} be saved in this table. All rows \textit{should} describe detectors that are referenced by Wavefront Sensors or Scoring Cameras.

\begin{table*}[htbp] \footnotesize
\caption{\label{tab:AOTDetectors}AOT\_DETECTORS fields.}
\centering
\begin{tabularx}{\linewidth}{>{\ttfamily}lllX}
\toprule
\normalfont{\textbf{Name}} & \textbf{Type} & \textbf{Unit} & \textbf{Description}\\
\midrule
UID                         &\textit{str}   &\textit{n.a.}  &Unique ID that identifies the detector.\tablefootmark{a,b}\\
TYPE                        &\textit{str}   &\textit{n.a.}  &Identifies the type of detector, such as \texttt{'CMOS'} or \texttt{'CCD'}.\\
SAMPLING\_TECHNIQUE         &\textit{str}   &\textit{n.a.}  &Identifies the sampling technique, for example \texttt{'Single Reset Read'}, \texttt{'Fowler'}, \texttt{'Double Correlated'}, \texttt{'Uncorrelated'}.\\
SHUTTER\_TYPE               &\textit{str}   &\textit{n.a.}  &Identifies the shutter type, typically \texttt{'Rolling'} or \texttt{'Global'}.\\
FLAT\_FIELD                 &\textit{str}   &\textit{n.a.}  &Image-reference. \\
READOUT\_NOISE              &\textit{flt}   &e\(^-\) s\(^{-1}\) pix\(^{-1}\)            &Digital detector read noise generated by its output amplifier in photo-electrons / second / pixel.\\
PIXEL\_INTENSITIES          &\textit{str}   &\textit{n.a.}  &Image-reference.\\
FIELD\_CENTRE\_X            &\textit{flt}   &\textit{pix}   &Defines the horizontal coordinate of the detector pixel on which the centre of the field is projected.\tablefootmark{d}\\
FIELD\_CENTRE\_Y            &\textit{flt}   &\textit{pix}   &Defines the vertical coordinate of the detector pixel on which the centre of the field is projected.\tablefootmark{d}\\
INTEGRATION\_TIME           &\textit{flt}   &s              &Duration in seconds a pixel integrates flux, independent of the detector reading scheme.\\
COADDS                      &\textit{int}   &count          &Number of frame co-additions.\\
DARK                        &\textit{str}   &\textit{n.a.}  &Image-reference.\\
WEIGHT\_MAP                 &\textit{str}   &\textit{n.a.}  &Image-reference.\\
QUANTUM\_EFFICIENCY         &\textit{flt}   &\textit{n.a.}  &A 0--1 scalar indicating the ability to convert a photon into a usable electron. Quoted at the central wavelength as an effective value.\\
PIXEL\_SCALE                &\textit{flt}   &rad pix\(^{-1}\) &Resolution in radians per detector pixel.\\
BINNING                     &\textit{int}   &count          &Integer value indicating the 2D pixel combination by the binning factor\\
BANDWIDTH                   &\textit{flt}   &m              &Width of the filter/bandpass of the optics+detector.\\
TRANSMISSION\_WAVELENGTH    &\textit{lst}   &m              &List of wavelengths that describe a transmission profile.\tablefootmark{c}\\
TRANSMISSION                &\textit{lst}   &\textit{n.a.}  &List of transmission percentages that describe a transmission profile.\tablefootmark{c}\\
SKY\_BACKGROUND             &\textit{str}   &\textit{n.a.}  &Image-reference.\\
GAIN                        &\textit{flt}   &e\(^-\)        &Scalar magnitude of detector signal amplification.\\
EXCESS\_NOISE               &\textit{flt}   &e\(^-\)        &Photon-noise gain factor (scalar) as a result of the electron-multiplied gain amplification in EMCCDs\\
FILTER                      &\textit{str}   &\textit{n.a.}  &Name of filter in use.\\
BAD\_PIXEL\_MAP             &\textit{str}   &\textit{n.a.}  &Image-reference.\\
DYNAMIC\_RANGE              &\textit{flt}   &dB             &Ratio of the maximum signal that can be integrated to the r.m.s. noise floor. If this ratio is R, then dynamic range in decibels is 20 log R.\\
READOUT\_RATE               &\textit{flt}   &pix s\(^{-1}\) &Inverse of the time required to digitize a single pixel.\\
FRAME\_RATE                 &\textit{flt}   &frame s\(^{-1}\)            &Inverse of the time needed for the detector to acquire an image and then completely read it out.\\
TRANSFORMATION\_MATRIX      &\textit{str}   &\textit{n.a.}  &Image-reference.\\
\bottomrule
\end{tabularx}
\tablefoot{\\
    \tablefoottext{a}{Strictly mandatory (value \textit{must not} be null).}
    \tablefoottext{b}{All entries in this field \textit{must} be unique.}
    \tablefoottext{c}{For each row in the table, entries in these fields \textit{must} have the same length, equal to the number of wavelengths at which transmission was calculated.}
    \tablefoottext{d}{A fractional value implies the centre is located in-between two pixels.}
}
\end{table*}

\subsubsection{AOT\_SCORING\_CAMERAS}
The fields of this binary table are described in Table~\ref{tab:AOTScoringCameras}. Some general information about each scoring camera in the system \textit{may} be provided here. Scoring cameras are able to reference the detector being used and an existing optical aberration.

\begin{table}[htbp] \footnotesize
\caption{\label{tab:AOTScoringCameras}AOT\_SCORING\_CAMERAS fields.}
\centering
\begin{tabularx}{\linewidth}{>{\ttfamily}lllX}
\toprule
\normalfont{\textbf{Name}} & \textbf{Type} & \textbf{Unit} & \textbf{Description}\\
\midrule
UID                 &\textit{str}   &\textit{n.a.}  &Unique ID that identifies the camera.\tablefootmark{a,b}\\
PUPIL\_MASK         &\textit{str}   &\textit{n.a.}  &Image-reference.\\
WAVELENGTH          &\textit{flt}   &m              &Observation wavelength \\
TRANSFORMATION\_MATRIX  &\textit{str}   &\textit{n.a.}      &Image-reference.\\
DETECTOR\_UID       &\textit{str}   &\textit{n.a.}  &AOT\_DETECTORS row-reference.\\
ABERRATION\_UID     &\textit{str}   &\textit{n.a.}  &AOT\_ABERRATIONS row-reference.\\
\bottomrule
\end{tabularx}
\tablefoot{\\
    \tablefoottext{a}{Strictly mandatory (value \textit{must not} be null).}
    \tablefoottext{b}{All entries in this field \textit{must} be unique.}
}
\end{table}

\subsubsection{AOT\_WAVEFRONT\_SENSORS}
The fields of this binary table are described in Table~\ref{tab:AOTWavefrontSensors}.

This table contains data for all wavefront sensors in the system, which \textit{may} be Shack-Hartmann or Pyramid. Wavefront sensors are able to reference the detector being used and an existing optical aberration. Wavefront sensors \textit{may} also define the non-common path aberration between them and the science detector. Rows \textit{must} reference the source that they are sensing. Each wavefront sensor \textit{may} be referenced by multiple controls loops that interact with it.

Both types of wavefront sensors \textit{may} provide secondary data through the tables \ref{tab:AOTWavefrontSensorsSH} or \ref{tab:AOTWavefrontSensorsP}.

\begin{figure*}[htbp]
    \centering
    \subfloat[Example \(80 \times 80\) \texttt{PIXEL\_INTENSITIES} image from the wavefront sensor's detector.]{
        \includegraphics[width=.3\linewidth]{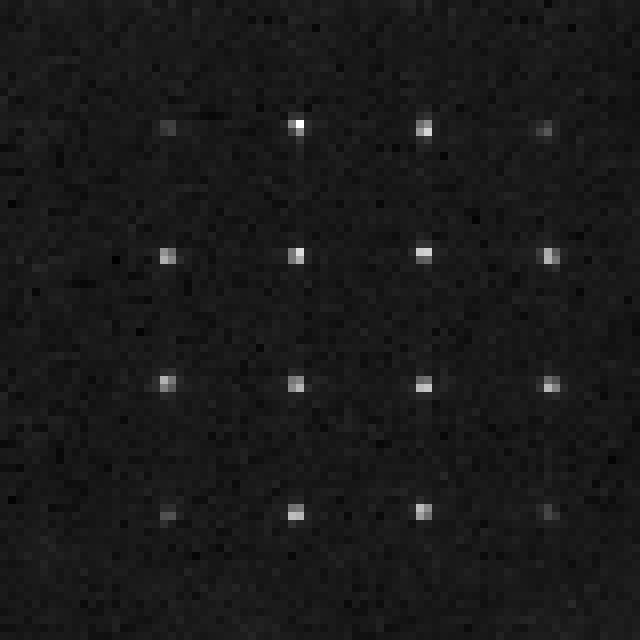}
        \label{subfig:pixels}
    }\hspace{1em}
    \subfloat[Example \(4 \times 4\) \texttt{SUBAPERTURE\_MASK} for a wavefront sensor with 12 valid subapertures. Cells with value \(-1\) indicate an invalid subaperture, while the others indicate the index of that subaperture in the sensor data.]{
        \includegraphics[width=.3\linewidth]{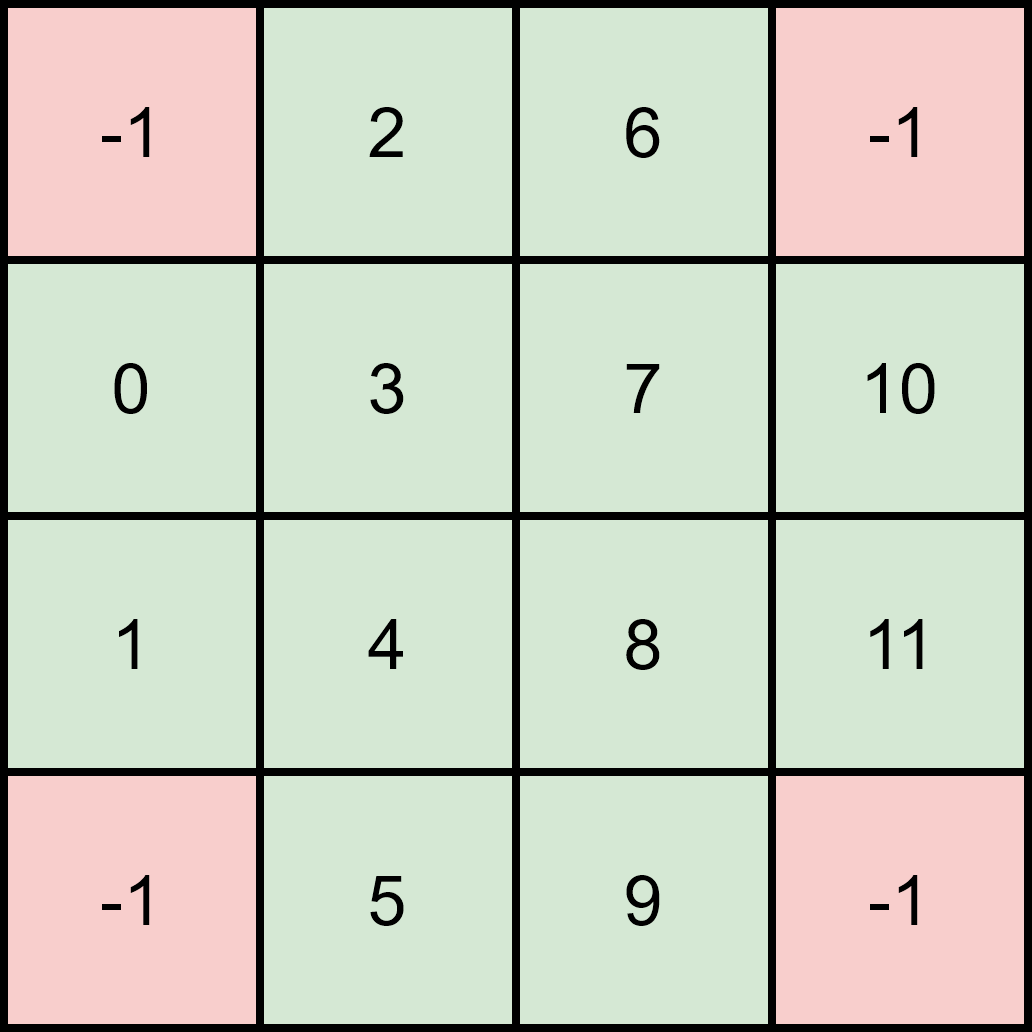}
        \label{subfig:grid}
    }\hspace{1em}
    \subfloat[\texttt{SUBAPERTURE\_MASK} superimposed over \texttt{PIXEL\_INTENSITIES}. In this case, \texttt{MASK\_X\_OFFSETS} and \texttt{MASK\_Y\_OFFSETS} have length one, indicating that the mask is superimposed once, and its lowermost positions in X and Y are 13 and 8, respectively. \texttt{SUBAPERTURE\_SIZE} defines each subaperture as corresponding to 16 pixels.]{
        \includegraphics[width=.3\linewidth]{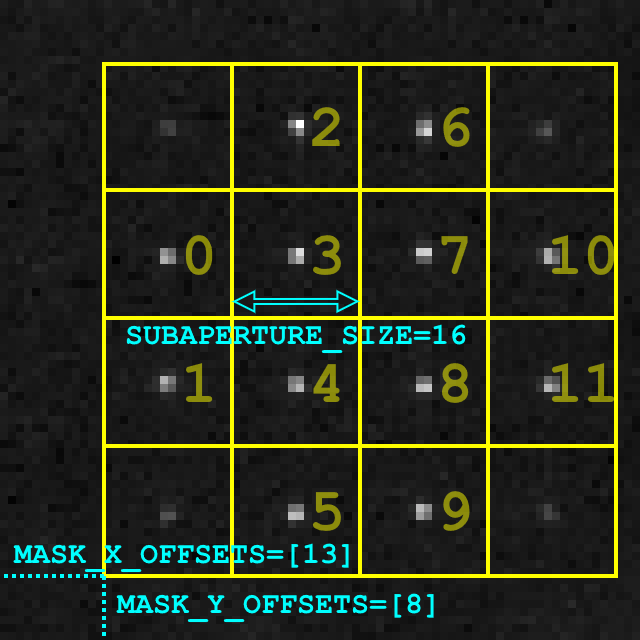}
        \label{subfig:imposed}
    }
    \caption{Diagram showing how the relationship between wavefront sensor subapertures and detector pixels can be deduced from \texttt{SUBAPERTURE\_MASK}, \texttt{MASK\_X\_OFFSETS}, \texttt{MASK\_Y\_OFFSETS} and \texttt{SUBAPERTURE\_SIZE}.}
    \label{fig:subaperturesdiagram}
\end{figure*}

\texttt{SUBAPERTURE\_MASK}, \texttt{MASK\_X\_OFFSETS}, \texttt{MASK\_Y\_OFFSETS} and \texttt{SUBAPERTURE\_SIZE} together describe the relationship between the detector's pixels and the wavefront sensor's subapertures. This mechanism is exemplified in Fig.~\ref{fig:subaperturesdiagram}.

\begin{table*}[htbp] \footnotesize
\caption{\label{tab:AOTWavefrontSensors}AOT\_WAVEFRONT\_SENSORS fields.}
\centering
\begin{tabularx}{\linewidth}{>{\ttfamily}lllX}
\toprule
\normalfont{\textbf{Name}} & \textbf{Type} & \textbf{Unit} & \textbf{Description}\\
\midrule
UID                         &\textit{str}   &\textit{n.a.}  &Unique ID that identifies the wavefront sensor.\tablefootmark{a,b}\\
TYPE                        &\textit{str}   &\textit{n.a.}  &Indicates the type of wavefront sensor (either \texttt{'Shack-Hartmann'} or \texttt{'Pyramid'}).\tablefootmark{a}\\
SOURCE\_UID                 &\textit{str}   &\textit{n.a.}  &AOT\_SOURCES row-reference. Indicates source being sensed.\tablefootmark{a}\\
DIMENSIONS                  &\textit{int}   &count  &Number of dimensions being measured by each subaperture. For \texttt{'Shack-Hartmann'} this \textit{must} be equal to 2 (horizontal and vertical offset). For \texttt{'Pyramid'} this \textit{must} be equal to either 2 (if the signals are also interpreted as horizontal and vertical offsets), 1 (if the subapertures overlap and are interpreted as a single signal) or the number of sides of the pyramid  (that is, \texttt{N\_SIDES} signals).\tablefootmark{a}\\
N\_VALID\_SUBAPERTURES      &\textit{int}   &count          & Number of valid subapertures (\textit{must} coincide with \texttt{SUBAPERTURE\_MASK} data).\tablefootmark{a}\\
MEASUREMENTS                &\textit{str}   &\textit{n.a.}  &Image-reference.\\
REF\_MEASUREMENTS           &\textit{str}   &\textit{n.a.}  &Image-reference.\\
SUBAPERTURE\_MASK           &\textit{str}   &\textit{n.a.}  &Image-reference.\\
MASK\_X\_OFFSETS            &\textit{lst}   &pix            &List of horizontal offsets in detector pixels. Each offset defines the lowest horizontal position occupied by the respective mask. \tablefootmark{c}\\
MASK\_Y\_OFFSETS            &\textit{lst}   &pix            &List of vertical offsets in detector pixels. Each offset defines the lowest vertical position occupied by the respective mask. \tablefootmark{c}\\
SUBAPERTURE\_SIZE           &\textit{flt}   &pix            &Size of each subaperture in detector pixels.\\
SUBAPERTURE\_INTENSITIES    &\textit{str}   &\textit{n.a.}  &Image-reference.\\
WAVELENGTH                  &\textit{flt}   &m              &Wavelength being sensed.\\
OPTICAL\_GAIN               &\textit{str}   &\textit{n.a.}  &Image-reference.\\
TRANSFORMATION\_MATRIX      &\textit{str}   &\textit{n.a.}  &Image-reference.\\
DETECTOR\_UID               &\textit{str}   &\textit{n.a.}  &AOT\_DETECTORS row-reference.\\
ABERRATION\_UID             &\textit{str}   &\textit{n.a.}  &AOT\_ABERRATIONS row-reference.\\
NCPA\_UID                   &\textit{str}   &\textit{n.a.}  &AOT\_ABERRATIONS row-reference, which describes the non-common path aberrations between the wavefront sensor and science detector.\\
\bottomrule
\end{tabularx}
\tablefoot{\\
    \tablefoottext{a}{Strictly mandatory (value \textit{must not} be null).}
    \tablefoottext{b}{All entries in this field \textit{must} be unique.}
    \tablefoottext{c}{For each row in the table, entries in these fields \textit{must} have the same length, equal to the number of masks projected on the detector pixels (there \textit{must} be one mask for \texttt{'Shack-Hartmann'} and as many as the number of pyramid sides for \texttt{'Pyramid'}).}
}
\end{table*}

\begin{table}[htbp] \footnotesize
\caption{\label{tab:AOTWavefrontSensorsSH}AOT\_WAVEFRONT\_SENSORS\_SHACK\_HARTMANN fields.}
\centering
\begin{tabularx}{\linewidth}{>{\ttfamily}lllX}
\toprule
\normalfont{\textbf{Name}} & \textbf{Type} & \textbf{Unit} & \textbf{Description}\\
\midrule
UID                         &\textit{str}       &\textit{n.a.}  &Unique ID that identifies the wavefront sensor.\tablefootmark{a,b}\\
CENTROIDING\_ALGORITHM      &\textit{str}       &\textit{n.a.}  &Name of the centroiding algorithm used.\\
CENTROID\_GAINS             &\textit{str}       &\textit{n.a.}  &Image-reference.\\
SPOT\_FWHM                  &\textit{str}       &\textit{n.a.}  &Image-reference.\\
\bottomrule
\end{tabularx}
\tablefoot{\\
    Secondary table exclusively for wavefront sensors of type \texttt{'Shack-Hartmann'}. If there are entries of this type in the mandatory table, there \textit{must} also exist an entry with the same \texttt{UID} in this table.\\
    \tablefoottext{a}{Strictly mandatory (value \textit{must not} be null).}
    \tablefoottext{b}{All entries in this field \textit{must} be unique.}
}
\end{table}

\begin{table}[htbp] \footnotesize
\caption{\label{tab:AOTWavefrontSensorsP}AOT\_WAVEFRONT\_SENSORS\_PYRAMID fields.}
\centering
\begin{tabularx}{\linewidth}{>{\ttfamily}lllX}
\toprule
\normalfont{\textbf{Name}} & \textbf{Type} & \textbf{Unit} & \textbf{Description}\\
\midrule
UID                         &\textit{str}       &\textit{n.a.}  &Unique ID that identifies the wavefront sensor.\tablefootmark{a,b}\\
N\_SIDES                    &\textit{int}       &count          & Number of pyramid sides (typically 4).\\
MODULATION                  &\textit{flt}       &m              & Modulation amplitude.\\
\bottomrule
\end{tabularx}
\tablefoot{\\
    Secondary table exclusively for wavefront sensors of type \texttt{'Pyramid'}. If there are entries of this type in the mandatory table, there \textit{must} also exist an entry with the same \texttt{UID} in this table.\\
    \tablefoottext{a}{Strictly mandatory (value \textit{must not} be null).}
    \tablefoottext{b}{All entries in this field \textit{must} be unique.}
}
\end{table}

\subsubsection{AOT\_WAVEFRONT\_CORRECTORS}
The fields of this binary table are described in Table~\ref{tab:AOTWavefrontCorrectors}.

This table contains data for all wavefront correctors in the system, which \textit{may} be Linear Stages (a correction that is applied in only one axis, usually focus), Tip-Tilt Mirrors (TTM, which correct in two axis, tip and tilt) or Deformable Mirrors (DM, which can correct multiple aberrations).

Rows in this table \textit{may} reference relevant optical aberrations that \textit{may} exist. They \textit{must} reference the telescope in which the corrector is installed. Each row \textit{may} be referenced by multiple system loops that interact with the corrector.

Deformable mirrors \textit{may} provide secondary data through Table~\ref{tab:AOTWavefrontCorrectorsDM}.

\begin{table*}[htbp] \footnotesize
\caption{\label{tab:AOTWavefrontCorrectors}AOT\_WAVEFRONT\_CORRECTORS fields.}
\centering
\begin{tabularx}{\linewidth}{>{\ttfamily}lllX}
\toprule
\normalfont{\textbf{Name}} & \textbf{Type} & \textbf{Unit} & \textbf{Description}\\
\midrule
UID                 &\textit{str}   &\textit{n.a.}  &Unique ID that identifies the wavefront corrector.\tablefootmark{a,b}\\
TYPE                &\textit{str}   &\textit{n.a.}  &Indicates the type of wavefront corrector (either \texttt{'Deformable Mirror'}, \texttt{'Tip-Tilt Mirror'} or \texttt{'Linear Stage'}).\tablefootmark{a}\\
TELESCOPE\_UID      &\textit{str}   &\textit{n.a.}  &AOT\_TELESCOPES row-reference. Indicates which telescope the corrector is installed on.\tablefootmark{a}\\
N\_VALID\_ACTUATORS &\textit{int}   &count          &Number of valid actuators in the corrector. This \textit{must} be 1 for \texttt{'Linear Stage'}, 2 for \texttt{'Tip-Tilt Mirror'}.\\
PUPIL\_MASK         &\textit{str}   &\textit{n.a.}  &Image-reference.\\
TFZ\_NUM            &\textit{lst}   &\textit{n.a.}  &List of numerators of the transfer function Z.\\
TFZ\_DEN            &\textit{lst}   &\textit{n.a.}  &List of denominators of the transfer function Z.\\
TRANSFORMATION\_MATRIX      &\textit{str}   &\textit{n.a.}      &Image-reference.\\
ABERRATION\_UID     &\textit{str}   &\textit{n.a.}  &AOT\_ABERRATIONS row-reference.\\
\bottomrule
\end{tabularx}
\tablefoot{\\
    \tablefoottext{a}{Strictly mandatory (value \textit{must not} be null).}
    \tablefoottext{b}{All entries in this field \textit{must} be unique.}
}
\end{table*}

\begin{table*}[htbp] \footnotesize
\caption{\label{tab:AOTWavefrontCorrectorsDM}AOT\_WAVEFRONT\_CORRECTORS\_DM fields.}
\centering
\begin{tabularx}{\linewidth}{>{\ttfamily}lllX}
\toprule
\normalfont{\textbf{Name}} & \textbf{Type} & \textbf{Unit} & \textbf{Description}\\
\midrule
UID                     &\textit{str}   &\textit{n.a.}  &Unique ID that identifies the wavefront corrector.\tablefootmark{a,b}\\
ACTUATORS\_X            &\textit{lst}   &m              &List of horizontal coordinates of the valid actuators of the DM.\tablefootmark{c}\\
ACTUATORS\_Y            &\textit{lst}   &m              &List of vertical coordinates of the valid actuators of the DM.\tablefootmark{c}\\
INFLUENCE\_FUNCTION     &\textit{str}   &\textit{n.a.}  &Image-reference.\\
STROKE                  &\textit{flt}   &m              &Maximum possible actuator displacement, measured as an excursion from a central null position.\\
\bottomrule
\end{tabularx}
\tablefoot{\\
    Secondary table exclusively for wavefront correctors of type \texttt{'Deformable Mirror'}. If there are entries of this type in the mandatory table, there \textit{must} also exist an entry with the same \texttt{UID} in this table.\\
    \tablefoottext{a}{Strictly mandatory (value \textit{must not} be null).}
    \tablefoottext{b}{All entries in this field \textit{must} be unique.}
    \tablefoottext{c}{For each row in the table, entries in these fields \textit{must} have the same length, equal to the number of valid actuators as defined by \texttt{N\_VALID\_ACTUATORS}.}
}
\end{table*}

\subsubsection{AOT\_LOOPS}
This binary table is described in Table~\ref{tab:AOTLoops}.

Each row in the table aggregates data related to one system loop. In this context, loops are understood as the conceptual mechanism through which the RTC periodically gathers data from a certain input, performs some calculation in those data, and outputs a set of commands to some sort of wavefront corrector.

Rows \textit{may} describe Control Loops, in which wavefront data (as sensed by a wavefront sensor) are taken as the input, from which the RTC calculates the necessary adjustments to a wavefront corrector (commands), based on interaction and control matrices. Alternatively, rows may describe Offload Loops, which take the set of commands received by one wavefront corrector as the input and, based on a offload matrix, output commands to another wavefront corrector (offloading the commands).

In the real-world, an AO loop \textit{may} command multiple wavefront correctors based on a single wavefront sensor, or even multiple wavefront sensors \textit{may} be used to command a single wavefront corrector (many-to-many relationships). While by design each row in this table can only describe a one-to-one relationship, AOT can still be used to describe many-to-many relationships. This is achieved by creating multiple conceptual loops that represent the same real-world loop, each having different inputs or outputs, as necessary to fully describe the loop \footnote[1]{For example, if a real-world AO loop uses a single wavefront sensor to command multiple wavefront correctors, we can have one control loop row for each wavefront corrector being commanded. These rows will all reference the same wavefront sensor as input, but their output wavefront corrector will differ, as will the control matrices and commands accordingly. The overhead for this is minimal, given that the duplicated data will be composed almost entirely of references.}

Given that control loops \textit{must} reference the wavefront sensor and wavefront corrector being used (which themselves \textit{must} reference a source and a telescope respectively), the data chain related to a control loop can be fully explored through references, allowing the user to study how these elements interact.

\begin{table*}[htbp] \footnotesize
\caption{\label{tab:AOTLoops}AOT\_LOOPS fields.}
\centering
\begin{tabularx}{\linewidth}{>{\ttfamily}lllX}
\toprule
\normalfont{\textbf{Name}} & \textbf{Type} & \textbf{Unit} & \textbf{Description}\\
\midrule
UID                         &\textit{str}   &\textit{n.a.}  &Unique ID that identifies the loop.\tablefootmark{a,b}\\
TYPE                        &\textit{str}   &\textit{n.a.}  &Indicates the type of loop (either \texttt{'Control Loop'} or \texttt{'Offload Loop'}).\tablefootmark{a}\\
COMMANDED\_UID              &\textit{str}   &\textit{n.a.}  &AOT\_WAVEFRONT\_CORRECTORS row-reference. This identifies the wavefront corrector that is being commanded by this loop.\tablefootmark{a}\\
TIME\_UID                   &\textit{str}   &\textit{n.a.}  &AOT\_TIME row-reference.\\
STATUS                      &\textit{str}   &\textit{n.a.}  &Values \textit{must} be \texttt{'Open'} or \texttt{'Closed'}, indicating whether the loop was opened or closed for the duration of the data collection.\\
COMMANDS                    &\textit{str}   &\textit{n.a.}  &Image-reference.\\
REF\_COMMANDS               &\textit{str}   &\textit{n.a.}  &Image-reference.\\
FRAMERATE                   &\textit{flt}   &Hz             &Frequency at which the loop operates.\\
DELAY                       &\textit{flt}   &frame          &Full AO loop delay, measured from the mid-point of the integration to the mid-point of the command applied. Includes the RTC latency and other communication delays.\\
TIME\_FILTER\_NUM           &\textit{str}   &\textit{n.a.}  &Image-reference.\\
TIME\_FILTER\_DEN           &\textit{str}   &\textit{n.a.}  &Image-reference.\\
\bottomrule
\end{tabularx}
\tablefoot{\\
    \tablefoottext{a}{Strictly mandatory (value \textit{must not} be null).}
    \tablefoottext{b}{All entries in this field \textit{must} be unique.}
}
\end{table*}

\begin{table*}[htbp] \footnotesize
\caption{\label{tab:AOTLoopsControl}AOT\_LOOPS\_CONTROL fields.}
\centering
\begin{tabularx}{\linewidth}{>{\ttfamily}lllX}
\toprule
\normalfont{\textbf{Name}} & \textbf{Type} & \textbf{Unit} & \textbf{Description}\\
\midrule
UID                         &\textit{str}   &\textit{n.a.}  &Unique ID that identifies the loop.\tablefootmark{a,b}\\
INPUT\_SENSOR\_UID          &\textit{str}   &\textit{n.a.}  &AOT\_WAVEFRONT\_SENSORS row-reference, which contains input data used to calculate commands, based on the \texttt{CONTROL\_MATRIX}.\tablefootmark{a}\\
MODES                       &\textit{str}   &\textit{n.a.}  &Image-reference.\\
MODAL\_COEFFICIENTS         &\textit{str}   &\textit{n.a.}  &Image-reference.\\
CONTROL\_MATRIX             &\textit{str}   &\textit{n.a.}  &Image-reference.\\
MEASUREMENTS\_TO\_MODES     &\textit{str}   &\textit{n.a.}  &Image-reference.\\
MODES\_TO\_COMMANDS         &\textit{str}   &\textit{n.a.}  &Image-reference.\\
INTERACTION\_MATRIX         &\textit{str}   &\textit{n.a.}  &Image-reference.\\
COMMANDS\_TO\_MODES         &\textit{str}   &\textit{n.a.}  &Image-reference.\\
MODES\_TO\_MEASUREMENTS     &\textit{str}   &\textit{n.a.}  &Image-reference.\\
RESIDUAL\_COMMANDS          &\textit{str}   &\textit{n.a.}  &Image-reference.\\
\bottomrule
\end{tabularx}
\tablefoot{\\
    Secondary table exclusively for loops of type \texttt{'Control Loop'}. If there are entries of this type in the mandatory table, there \textit{must} also exist an entry with the same \texttt{UID} in this table.\\
    \tablefoottext{a}{Strictly mandatory (value \textit{must not} be null).}
    \tablefoottext{b}{All entries in this field \textit{must} be unique.}
}
\end{table*}

\begin{table*}[htbp] \footnotesize
\caption{\label{tab:AOTLoopsOffload}AOT\_LOOPS\_OFFLOAD fields.}
\centering
\begin{tabularx}{\linewidth}{>{\ttfamily}lllX}
\toprule
\normalfont{\textbf{Name}} & \textbf{Type} & \textbf{Unit} & \textbf{Description}\\
\midrule
UID                         &\textit{str}   &\textit{n.a.}  &Unique ID that identifies the loop.\tablefootmark{a,b}\\
INPUT\_CORRECTOR\_UID       &\textit{str}   &\textit{n.a.}  &AOT\_WAVEFRONT\_CORRECTORS row-reference, from which commands are offloaded, based on the \texttt{OFFLOAD\_MATRIX}.\tablefootmark{a}\\
OFFLOAD\_MATRIX             &\textit{str}   &\textit{n.a.}  &Image-reference.\\
\bottomrule
\end{tabularx}
\tablefoot{\\
    Secondary table exclusively for loops of type \texttt{'Offload Loop'}. If there are entries of this type in the mandatory table, there \textit{must} also exist an entry with the same \texttt{UID} in this table.\\
    \tablefoottext{a}{Strictly mandatory (value \textit{must not} be null).}
    \tablefoottext{b}{All entries in this field \textit{must} be unique.}
}
\end{table*}

\subsection{Images} \label{ss:images}
In AOT, multidimensional data is always stored in images extensions that \textit{may} be referenced. These image extensions \textit{may} be contained in the AOT file itself (after all the binary tables), or in separate FITS files (see Appendix~\ref{sapp:crossreferencing}).

There are 35 different categories of images that \textit{may} be referenced, as seen in Table~\ref{tab:Images}. For each image category, the table specifies the data type, their expected dimensions, their units and a description that characterises the data. In a single AOT file there \textit{may} be any number of images of the same category and the same image \textit{may} be referenced by multiple entries. This means that, in cases where different table rows need to reference the exact same multidimensional data, it is highly \textit{recommended} that only a single image extension is created for that data, avoiding data duplication. Images that are never referenced in the file \textit{should not} be included.

As mentioned in Appendix~\ref{sapp:crossreferencing}, each image extension \textit{must} have a unique non-null name (as set through \texttt{EXTNAME}), which is defined by the user. It is \textit{recommended} that images are named in a way that allows the file to be self-explanatory. Image extensions \textit{must} respect the data type (as set through \texttt{BITPIX}) and dimensions (as set through \texttt{NAXIS}) specified in the table. If the unit of measurement of the data in a certain image is not specified (through \texttt{BUNIT}), it is assumed it follows the table specification; if it is different from the specification, the unit \textit{must} be specified (through \texttt{BUNIT}). Whenever the unit of an image does not follow the specification, it is \textit{recommended} that it is kept consistent with other images of the same category, as well as related image categories. Users are free to further describe the contents of an image extension through the addition of any user-defined metadata keywords in the header, as long as the FITS specification is respected.

All images \textit{may} have a \(t\) dimension (defined after all the dimensions specified in Table~\ref{tab:Images}); in such cases, the data in that image is understood to be time-varying. It is \textit{recommended} that time-varying images contain a \texttt{TIME\_UID} keyword in their header, whose value is a AOT\_TIME row-reference. This row \textit{must} have time data of the same length as the \(t\) dimension, allowing the user to relate the data to the specific instants it applies to. If an image is not time-varying (e.g. it applies to the entire recording, or it describes average values over the relevant timeframe rather than the values at a specific instant), the \(t\) dimension \textit{may} be omitted and the \texttt{TIME\_UID} keyword \textit{must} not be present. Although technically supported, we recommend not saving time-varying images containing AOT\_TIME row-references externally (for example, to be used as external image-references, as described in Appendix~\ref{sapp:crossreferencing}); this is because the row-reference has no meaning outside the particular AOT file. Overall, external image-references are better suited to avoid data duplication for images which remain unchanged throughout multiple observations (that is, typically images that contain non-time-varying data).

{\footnotesize
\onecolumn
\begin{xltabular}{\linewidth}{>{\ttfamily}llllX} 
\caption{\label{tab:Images}List of possible image categories.}\\
\toprule
\normalfont{\textbf{Image category}} & \textbf{Type} & \textbf{Dimensions (\(\times t\))} & \textbf{Unit} & \textbf{Description}\\
\midrule
\multicolumn{5}{c}{\normalfont{\textit{Referenceable in multiple tables}}}\\
TRANSFORMATION\_MATRIX      & \textit{flt}  & \(3 \times 3\)           & \textit{n.a.}     & Matrix that defines 2-dimensional affine transformations using homogeneous coordinates. For more detail see Appendix~\ref{sapp:geometry}.\\
PUPIL\_MASK                 & \textit{int}  & \(w \times h\)                    & \textit{n.a.}     & Binary image that describes the shape of the pupil. A 1 indicates the presence of the pupil, while a 0 indicates the opposite.\\
MODES                       & \textit{flt}  & \(w \times h \times m\)           & \textit{n.a.}     & Set of \(m\) different \(w \times h\) arrays, each representing the orthonormal basis of the corresponding mode.\\
\midrule
\multicolumn{5}{c}{\normalfont{\textit{AOT\_ATMOSPHERIC\_PARAMETERS}}}\\
LAYERS\_WEIGHT              & \textit{flt}  & \(l\)                    & \textit{n.a.}     & Fractional weight of each \(l\) turbulence layer (sums to 1).\\
LAYERS\_HEIGHT              & \textit{flt}  & \(l\)                    & m                 & Height above observatory at zenith of each \(l\) turbulence layer.\\
LAYERS\_L0                  & \textit{flt}  & \(l\)                    & m                 & Outer scale of turbulence at reference wavelength at zenith of each \(l\) turbulence layer.\\
LAYERS\_WIND\_SPEED         & \textit{flt}  & \(l\)                    & ms\(^{-1}\)       & Wind speed of each \(l\) turbulence layer.\\
LAYERS\_WIND\_DIRECTION     & \textit{flt}  & \(l\)                    & deg               & Wind direction of each \(l\) turbulence layer, with 0\degree\ being North, increasing eastward.\\
\midrule
\multicolumn{5}{c}{\normalfont{\textit{AOT\_ABERRATIONS}}}\\
COEFFICIENTS                & \textit{flt}  & \(n \times m\)                    & \textit{u.d.}     & Set of \(m\) coefficients (one for each of the orthonormal basis of modes) for each \(n\) pupil offset.\\
\midrule
\multicolumn{5}{c}{\normalfont{\textit{AOT\_SOURCES\_SODIUM\_LGS}}}\\
PROFILE                     & \textit{flt}  & \(l\)                    & \textit{n.a.}     & Normalised LGS profile (each set of \(l\) layers \(\sum = 1\)) at zenith.\\
\midrule
\multicolumn{5}{c}{\normalfont{\textit{AOT\_DETECTORS}}}\\
FLAT\_FIELD                 & \textit{flt}  & \(w \times h\)                    & \textit{n.a.}     & Inverse of the detector pixel-to-pixel sensitivity.\\
PIXEL\_INTENSITIES          & \textit{flt}  & \(w \times h\)           & ADU               & Intensity detected in each pixel, for each data frame. This is an image spanning \(w\) pixels horizontally and \(h\) pixels vertically.\\
DARK                        & \textit{flt}  & \(w \times h\)                    & ADU               & Intensity detected in each pixel when there is no light being observed. This is an image spanning \(w\) pixels horizontally and \(h\) pixels vertically.\\
WEIGHT\_MAP                 & \textit{flt}  & \(w \times h\)                    & \textit{n.a.}     & Pixel weight map, where each detector pixel is associated with a number that represents its relative value. Summing up to 1.\\
SKY\_BACKGROUND             & \textit{flt}  & \(w \times h\)                    & ADU               & Detector pixel intensities from a source-less direction in the sky.\\
BAD\_PIXEL\_MAP             & \textit{int}  & \(w \times h\)                    & \textit{n.a.}     & Binary image which identifies the bad pixels, that is, pixels that should be ignored for the purpose of the analysis. Pixels identified with 1 are considered bad, while 0 is considered normal.\\
\midrule
\multicolumn{5}{c}{\normalfont{\textit{AOT\_WAVEFRONT\_SENSORS}}}\\
MEASUREMENTS                & \textit{flt}  & \(s_v \times d\)         & \textit{u.d.}     & Wavefront measurements from the wavefront sensor. Each of its \(s_v\) subapertures is able to measure in \(d\) dimensions. \\
REF\_MEASUREMENTS           & \textit{flt}  & \(s_v \times d\)                  & \textit{u.d.}     & Reference measurements.\\
SUBAPERTURE\_MASK           & \textit{int}  & \(s \times s\)                    & \textit{n.a.}     & Representation of the subaperture grid, where the cells corresponding to invalid subapertures are marked as \(-1\) and the cells corresponding to valid subapertures contain their respective index in the sequence of valid subapertures (using zero-based numbering, that is, the initial element is assigned the index 0).\\
SUBAPERTURE\_INTENSITIES    & \textit{flt}  & \(s_v\)                  & ADU               & Detected average intensity (flux) of each of the \(s_v\) valid subapertures.\\
OPTICAL\_GAIN               & \textit{flt}  & \(1\)                             & \textit{n.a.}     & Wavefront sensor optical gain.\\
\midrule
\multicolumn{5}{c}{\normalfont{\textit{AOT\_WAVEFRONT\_SENSORS\_SHACK\_HARTMANN}}}\\
CENTROID\_GAINS             & \textit{flt}  & \(s_v \times d\)                  & \textit{n.a.}     & Centroid gain factors for each of \(s_v\) subapertures and \(d\) dimensions.\\
SPOT\_FWHM                  & \textit{flt}  & \(s_v \times d\)                  & arcsec            & Spot full width half maximum for each of \(s_v\) subapertures and  \(d\) dimensions.\\
\midrule
\multicolumn{5}{c}{\normalfont{\textit{AOT\_WAVEFRONT\_CORRECTORS\_DM}}}\\
INFLUENCE\_FUNCTION         & \textit{flt}  & \(w \times h \times a_v\)         & m                 & A set of 2D images, one for each valid actuator, where each image represents the displacement of the surface of the deformable mirror after poking the respective actuator. \\
\midrule
\multicolumn{5}{c}{\normalfont{\textit{AOT\_LOOPS}}}\\
COMMANDS                    & \textit{flt}  & \(a_v\)                  & m                 & Commands sent to the associated wavefront corrector. Contains the values sent to each of the \(a_v\) valid actuators of a certain wavefront corrector.\\
REF\_COMMANDS               & \textit{flt}  & \(a_v\)                           & m                 & Reference offset commands for each of \(a_v\) actuators.\\
TIME\_FILTER\_NUM           & \textit{flt}  & \(i \times m\)                    & \textit{n.a.}     & One set of time filter numerators per mode being described. If \(m=1\), it is assumed to be applicable to all modes. The first numerator is the loop gain. \\
TIME\_FILTER\_DEN           & \textit{flt}  & \(j \times m\)                    & \textit{n.a.}     & One set of time filter denominators per mode being described. If \(m=1\), it is assumed to be applicable to all modes. Uses standard transfer function \(sys(z) = \sum_{i=0}^{N-1} b_i z^i/\sum_{j=0}^N a_j z^j\).\\
\midrule
\multicolumn{5}{c}{\normalfont{\textit{AOT\_LOOPS\_CONTROL}}}\\
MODAL\_COEFFICIENTS         & \textit{flt}  & \(m \times t\)                    & \textit{u.d.}     & Coefficients respective to each of the \(m\) modes being corrected on the wavefront corrector.\\
CONTROL\_MATRIX             & \textit{flt}  & \(s_v \times d \times a_v\)       & \textit{u.d.}     & Linear relationship between the wavefront sensor measurements (\(s_v \times d\)) and the corrector commands (\(a_v\)).\\
MEASUREMENTS\_TO\_MODES     & \textit{flt}  & \(s_v \times d \times m\)         & \textit{u.d.}     & Linear relationship between the wavefront sensor measurements (\(s_v \times d\)) and the modes to be corrected (\(m\)).\\
MODES\_TO\_COMMANDS         & \textit{flt}  & \(m \times a_v\)                  & \textit{u.d.}     & Linear relationship between the modes (\(m\)) to be corrected and the corrector commands (\(a_v\)).\\
INTERACTION\_MATRIX         & \textit{flt}  & \(a_v \times d \times s_v\)       & \textit{u.d.}     & Represents the measurements of the wavefront sensor (\(d \times s_v\)) in response to each actuator of the wavefront corrector (\(a_v\)).\\
COMMANDS\_TO\_MODES         & \textit{flt}  & \(a_v \times m\)                  & \textit{u.d.}     & Represents the modal response (\(m\)) to each actuator of the wavefront corrector (\(a_v\)).\\
MODES\_TO\_MEASUREMENTS     & \textit{flt}  & \(m \times d \times s_v\)         & \textit{u.d.}     & Linear relationship between the modal response (\(m\)) and the wavefront sensor measurements (\(d \times s_v\)).\\
RESIDUAL\_COMMANDS          & \textit{flt}  & \(a_v\)                           & \textit{u.d.}     & Reconstructed corrector commands before time filtering.\\
\midrule
\multicolumn{5}{c}{\normalfont{\textit{AOT\_LOOPS\_OFFLOAD}}}\\
OFFLOAD\_MATRIX             & \textit{flt}  & \(a1_v \times a2_v\)              & \textit{u.d.}     & Linear relationship between the commands from one actuator (\(a1\)) and the corresponding commands that get offloaded to another actuator (\(a2\)).\\
\bottomrule
\end{xltabular}
\tablefoot{\\
    The unit "\textit{n.a.}" stands for "not applicable", meaning that the data is a dimensionless quantity, such as countable quantities, ratios and proportions. The unit "\textit{u.d.}" stands for "user defined" meaning that the user is free to pick a unit as long as it is consistent with the rest of the data.\\
    All images in this table \textit{may} vary over time. In such cases, the images \textit{must} have an extra \(t\) dimension, which \textit{must} defined after all the other specified dimensions. It is \textit{recommended} that images with a \(t\) contain a \texttt{TIME\_UID} keyword in their header, whose value is a AOT\_TIME row-reference with time data of \(t\) length. The \(t\) dimension may be omitted if it is of size 1.\\ 
    \(w\) and \(h\) are (respectively) the width and height of the image in pixels, \(s_v\) is the number of valid subapertures (ordered as defined via \texttt{SUBAPERTURE\_MASK}), \(d\) is the number of dimensions being measured by the wavefront sensor, \(a_v\) is the number of valid actuators, \(m\) is the number of modes being controlled.\\
    If an image of data type \textit{flt} can be accurately described with a \textit{int} type, users \textit{may} opt to do so. However, this does not apply to the inverse.
}
\twocolumn
}

\subsection{Extra data}
While there are no theoretical obstacles to creating a FITS file that contains extra HDUs, keywords or fields on top of the ones that were already explicitly defined in this specification, tools specially designed to support AOT files \textit{may} not be expecting such non-standard data. Consequentially, these extra data \textit{may} be ignored by supporting tools or, in a worst case scenario, cause unexpected behaviour. In order to ensure maximal compatibility, it is highly \textit{recommended} that such arbitrary modifications are avoided, particularly in files that are meant to be shared with third-parties. If adding some extra metadata is absolutely necessary (for example, to ensure compatibility with existing archives or pipelines), it is \textit{recommended} that these additions are limited to the primary header.

\section{Supporting tools} \label{sec:tools}
Given that the usefulness of a format of this kind is largely predicated on its wide adoption, reducing any existing barriers to adoption is of paramount importance. One of the ways user-friendliness can be improved is by providing the community with a set of supporting tools.

In general terms, an AOT supporting tool should provide functionalities that help users read, edit and write AOT files, while abstracting them from format specification details (such as layout of the file, table naming, mandatory keywords and the cross-referencing mechanisms). They may also provide provide translation mechanism for interacting with existing non-AOT telemetry datasets. Since AOT files potentially contain large amounts of data, it is recommended that tools avoid loading all data to memory (and instead load it as it becomes necessary) and that special care is taken to avoid unnecessary data duplication (particularly in image extensions).

The language-agnostic data model for AOT described in Sec.~\ref{ss:data_model} enables the abstractions described above while still allowing access to all relevant data. In a previous paper \citep{towards_SPIE_2022}, we discussed the development of \textit{aotpy} (Adaptive Optics Telemetry for Python), an object-oriented Python \citep{guido2009python} package based on the AOT data model, which was designed with the considerations described above; we will provide an overview of this package in the following section.

\subsection{aotpy} \label{ss:aotpy}
\textit{aotpy} is an implementation of the AOT data model along with other functionalities that improve the ease-of-use of the AOT format. In \textit{aotpy} a recording of AO telemetry data can be described by one Python object (named \textit{AOSystem}) which contains all the data and relationships between parts of the system (see data model in Fig.~\ref{fig:umlclassdiagram}).

\begin{figure}[htbp]
   \begin{center}
   \includegraphics[width=\columnwidth]{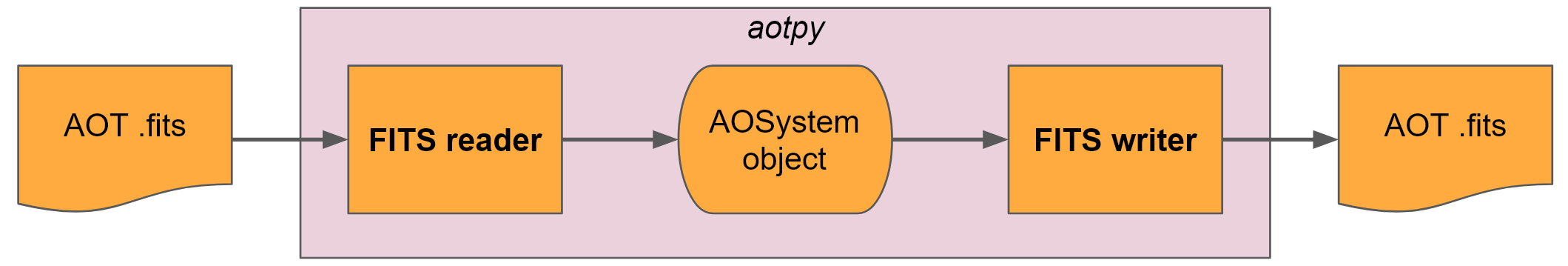}
   \end{center}
   \caption{\label{fig:readwrite} 
    Diagram that exemplifies how "readers" and "writers" work in \textit{aotpy}. The starting and final "AOT .fits" files are functionally equivalent. The \textit{AOSystem} object is completely agnostic from the specific file format in use, meaning that it would be possible to read a FITS file and write an equivalent file in another file format, as long as such a writer was implemented.}
\end{figure}

Using \textit{aotpy}, an AOT file can be read into a set of related objects (as defined in the data model) that enable the user to interact with the data in a generalised manner. These objects can also be automatically written back into an AOT file, completely abstracting the user from file handling details. These operations are performed by "writers" and "readers", such that one writer-reader pair must exist per each file format supported. While currently only the FITS format is implemented, the package is designed in a modular way that would allow for a simple expansion to further formats. A diagram of the writing and reading process is shown in Fig.~\ref{fig:readwrite}.

\begin{figure}[htbp]
   \begin{center}
   \includegraphics[width=\columnwidth]{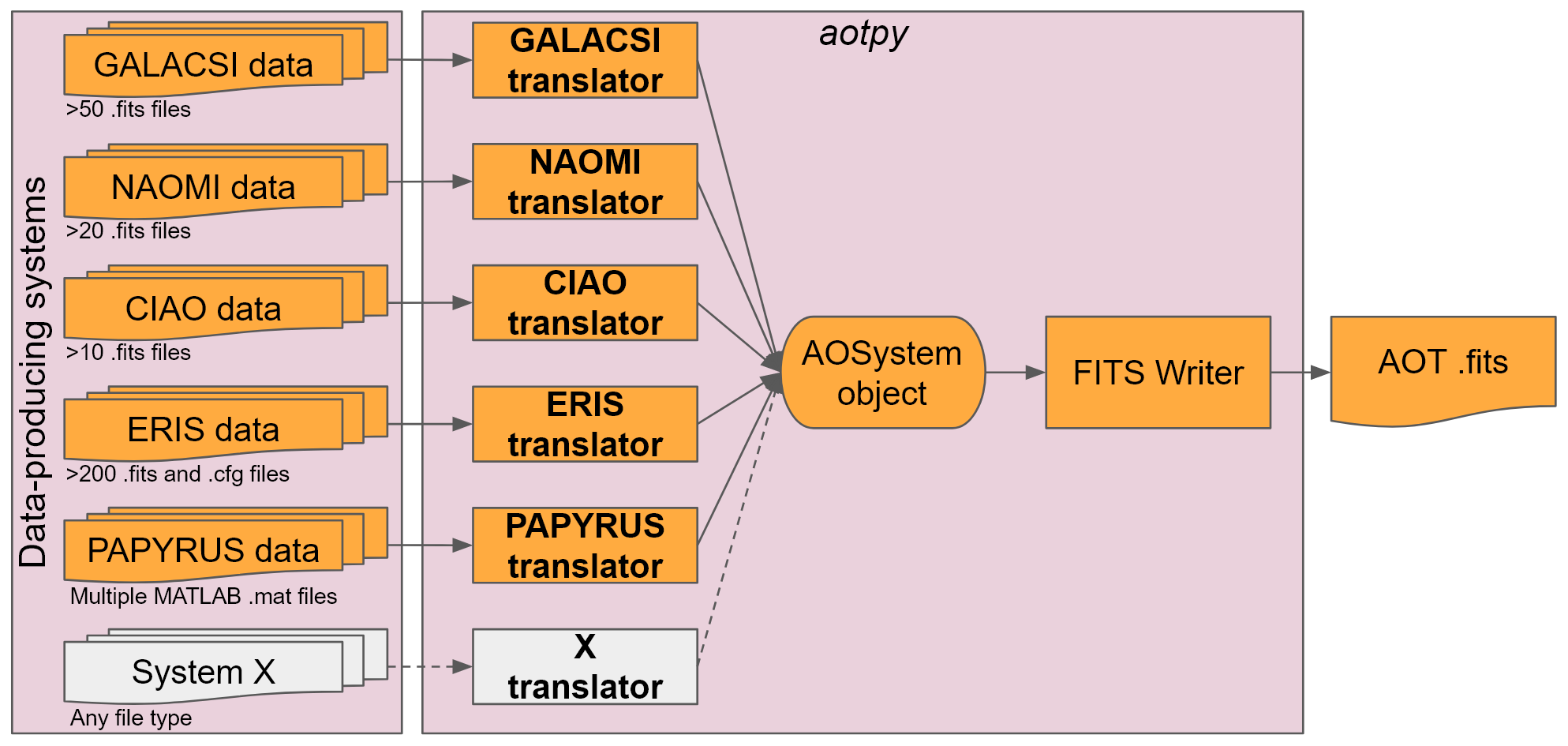}
   \end{center}
   \caption{\label{fig:translation} 
    Diagram that exemplifies how "translators" work in \textit{aotpy}. The files to the left are non-standard and may use any file format (or even multiple formats). Each system requires a specific translator that is able to interpret data from their respective files and fill the corresponding fields of the AOSystem object. This resulting object is completely agnostic from the system where the data originated from, and it can then be used as any other \textit{aotpy} object. In this example, it is written back into a single AOT .fits file using the FITS writer.}
\end{figure}

On top of the file handling functionalities, \textit{aotpy} is also able to translate some existing non-standard data files into a standard \textit{aotpy} object. This allows the user to handle data from systems that do not follow the AOT specification using \textit{aotpy}, which is achieved by providing one translation script ("translator") for each supported system. These scripts are able to read and interpret the files produced by the respective system, and convert that data into an \textit{AOSystem} object that describes the recording. This object is system-agnostic and can be treated as any other such \textit{AOSystem} object, meaning that it can, among other things, be written as a standard AOT FITS file. Currently, we have implemented proof-of-concept translators for ESO's GALACSI \citep{STUIK2006618}, NAOMI \citep{10.1117/12.2054730}, CIAO \citep{10.1117/12.926558} and ERIS \citep{10.1117/12.2234001}, as well as ALPAO's PAPYRUS \citep{10.1117/12.2597170}. A diagram exemplifying the translation process can be seen in Fig.~\ref{fig:translation}.

We have made \textit{aotpy} publicly available on PyPI (Python Package Index) and its source code is also available on GitHub and Zenodo \citep{aotpyGitHub, aotpyZenodo}\footnote{Accessible via \url{https://github.com/STAR-PORT/aotpy} and \url{https://zenodo.org/doi/10.5281/zenodo.8187229}, respectively.}.

\section{Demonstration and data availability}
In the interest of allowing users to have an immediate experience with AOT files, we made available some proof-of-concept files. Using \textit{aotpy}, we translated GALACSI, CIAO, NAOMI and ERIS data gathered under ESO's program IDs 60.A-9278(B), 60.A-9278(C), 60.A-9278(D) and 60.A-9278(E) respectively, and published the translated files on the ESO archive. These are publicly accessible and searchable under the respective program IDs\footnote{Queries on the ESO archive can be performed at \url{https://archive.eso.org/eso/eso_archive_main.html}. The "Program ID" textbox can be used to specify the instrument.}. A small dataset has also been made available on Zenodo \citep{dataZenodo}\footnote{Accessible at \url{https://zenodo.org/doi/10.5281/zenodo.8187229}.}; this dataset contains PAPYRUS data, along with a subset of the data also shared through the ESO archive. This demonstration shows the adaptability of the format to vastly different systems, including different telescopes, different AO modes, different wavefront sensor types and different control schemes.

The translation of the original non-AOT datasets into these proof-of-concept followed the process described in Sec.~\ref{ss:aotpy}; for example, converting ERIS data contained in a folder located at "path/to/folder" into a single "example.fits" AOT file can be done with the following code (as long as the \textit{aotpy} package is installed): 

\begin{minted}{python}
from aotpy.translators import ERISTranslator
sys = ERISTranslator('path/to/folder').translate()
sys.write_to_file('example.fits')
\end{minted}

In \cite{turbulenceNAOMI}, NAOMI AO telemetry data (made available by this project) is used to estimate integrated turbulence parameters, fully demonstrating the suitability of AOT for such use-cases.

The PSF-R use case is still in development and a full demonstration is not available at this time. We are supporting the developments made under the ELT working group for PSF-R\footnote{See \url{https://elt.eso.org/about/workinggroups/}.} in all its dimensions, namely with demonstrations on the VLT with ERIS in preparation for the full-blown deployment on the ELT. 

\section{Summary and Outlook}

In this paper, we highlight the immense scientific potential within the field of AO research. To harness this potential, we address the challenge of limited end-user access to pertinent data. Despite a growing interest among data-producing systems to share their valuable data, the community has yet to establish a consensus on the most effective approach. To overcome this challenge, we introduce AOT, an innovative data exchange format that has been collaboratively developed within the community. AOT is designed to accommodate a wide range of system configurations, empowering us to unlock the full potential of AO research.

AOT is focused on supporting scientifically important use-cases through a set of design principles that address the needs of the community. It is a FITS-based format, on top of which we have developed a set of conventions and structures that enable users to have access to relevant data in a standardised manner. This format is fully specified in order to avoid ambiguity and allow for interchangeability.

We discuss the need to provide the community with supporting tools that ease the adoption of the format. We demonstrate their use and provide public datasets.

\subsection{Future Work}
This first public version of AOT aims to get a significant foothold in the AO space, while acknowledging that future expansions may be necessary, as the field progresses and community needs change. Therefore, we are committed to providing continuous support and improvements in the foreseeable future.

Looking ahead, our foremost objectives include:
\begin{itemize}
    \item Continue developing and improving supporting tools and their documentation. We are currently in the early stages of developing a graphical user interface for AOT data, with the goal of providing a dashboard that allows users to perform exploratory data analysis interactively. In conjunction with the ESO Archive team, we are also developing new user-facing query forms that improve access to AOT data in the archive.
    \item Further promoting and discussing the format with both end-users and data-producing teams. We have initiated talks with other instrument teams (including the teams responsible for Keck AO \citep{Wizinowich_2000}, LBT AO \citep{Skemer_2012}, NFIRAOS \citep{10.1117/12.617849}, HARMONI \citep{10.1117/12.857445}, MICADO \citep{10.1117/12.856379} and METIS \citep{10.1117/12.789241}) to explore the possibility of also providing translation scripts for those systems; this diversification would further exemplify AOT's adaptability to different AO configurations. We are also exploring how AOT/\textit{aotpy} can be integrated into data processing pipelines currently in use by AO instruments.
    \item Exploring and demonstrating the usefulness of AOT in additional scenarios such as AIT phases, ELT-class telescopes, High-Contrast Imaging and Interferometry.
    
\end{itemize}
All in all, in the coming phases, the pivotal focus will be on establishing and maintaining enduring community support and funding for the AOT software package, ensuring its continuous development to meet the evolving needs of future scientific instruments and observatories. Transitioning from the current proof-of-concept phase to a fully operational setting will be instrumental in realising the adoption of AOT by the next generation of state-of-the-art instruments, underscoring the essential role of sustained financial and community backing in this endeavour.

\appendix
\section{Design principles} \label{app:design_principles}
In order to guide the necessary decision-making for designing the format, we deem that a set of design principles must be defined. These principles essentially illustrate the characteristics of a format that would best meet the goals of this format. Specifically, this idealised format would be:
\begin{itemize}
    \item Shareable: exchanging data between data-producing systems and their end-users must be a straightforward process.
    \item Unambiguous: ensure that data shared can only be interpreted in one well-defined manner, that is consistent across all systems.
    \item Generalised: structure the format in a way that allows the end-user to access data abstracted from observatory or instrument details.
    \item Flexible: support as many AO systems as possible, by being sufficiently adaptable to their differences and the irregularities of real-life telemetry.
    \item Complete: ensure that all relevant information can be packaged in the format.
    \item User-friendly: adoption barriers must be reduced by providing accessible tools and documentation.
    \item Expansible: allow for future expansions to meet user needs and advancements in the AO field.
\end{itemize}

In sum, the format structure and data access should be the exact same across various systems and configurations, while the data itself should be trivial to access and understand. Given that some of these design principles can conflict with each-other, in practice it is not possible to perfectly achieve all of them. Therefore, some design decisions have to be analysed as compromises between different principles. In general, we consider these principles should be prioritised in the listed order.

\section{FITS overview} \label{app:fits_overview}
The Flexible Image Transport System (FITS) is a data format first proposed in 1979 aimed at standardising the interchange of astronomical images and other digital arrays on magnetic tapes \citep{wells_fits-flexible_1979}. The format is now understood as being independent of physical medium and has since received multiple revisions that accommodate further features \citep{hanisch_definition_1993, hanisch_definition_2001, hanisch_definition_2005, pence_definition_2010, chiappetti_definition_2018}. 

In order to fully describe AOT we will need to reference many FITS concepts. Below, we provide an overview of the FITS format as a quick-reference guide, but the most recent FITS specification \citep[at the time of writing]{chiappetti_definition_2018} should be referred to for the full specification.

In general terms, a FITS file is a sequence of \textit{Header Data Units} (HDUs), with each HDU being composed of two parts: the \textit{header} and the \textit{data array}. The header is a series of ASCII text \textit{keywords} and their respective \textit{values}, that are typically used to describe the data array contained in that HDU (essentially, metadata). While there is a set of reserved keywords whose meaning is defined in the FITS specification (some of which are mandatory), any strings \textit{may} be used as keywords (as long as they are up to 8 characters in length). The data array itself follows the header, and it can be formatted in multiple ways depending on its specific type.

The first HDU in a file is known as the \textit{primary HDU} and it contains the \textit{primary header}, which is followed by either \textit{image} data or no data array at all. The primary HDU \textit{may} be followed by any number of HDUs, which are known as \textit{extensions}. Unlike the primary HDU, extensions \textit{may} contain data types other than images. The data type in use is identified by the value of the \texttt{XTENSION} keyword, which \textit{should} be one of the values approved by the IAU FITS Working Group \citep[Appendix~F]{chiappetti_definition_2018}. AOT only makes use of the \textit{Binary Table} (\texttt{BINTABLE}) and \textit{Image} (\texttt{IMAGE\textvisiblespace\textvisiblespace\textvisiblespace}) extensions. Sections \ref{ss:bintables} and \ref{ss:images} describe how these extensions are used in AOT, respectively.

Despite the name, Image extensions do not necessarily define an image; instead they \textit{may} contain any multidimensional homogeneous array (up to 999 dimensions). Each array is composed of either characters, integers or floating point values, for which metadata such as units of measurement \textit{may} be defined (in the corresponding header).

Binary Table extensions are tables composed of up to 999 ordered columns (known as \textit{fields}) and as many ordered rows as necessary. Each field has a case-insensitive name (represented in upper-case) and a certain data type that applies to all of its \textit{entries} (there is one entry per row). Other metadata about the field \textit{may} also be \textit{optionally} defined, such as units of measurement. All these column metadata are described using keywords of the form \texttt{TXXXX}\(n\) , where \texttt{TXXXX} represents one of an open set of mandatory, reserved or user-defined keywords up to five characters in length, for example: \texttt{TFORM} (mandatory), \texttt{TUNIT} (reserved), \texttt{TCOMM} (user-defined). The \(n\) part is an integer between 1 and 999 which indicates the index of the column that the keyword in question refers to.

FITS provides no limitation on the total number of HDUs nor on the the total size of the FITS file \citep[Sec.~3.1]{chiappetti_definition_2018}. Therefore, in practical terms the file size is solely restricted by the limits of the device's filesystem itself, or of any tools handling the data. We expect that, in real-world scenarios, this limitation is unlikely to be significant when using modern 64-bit based filesystems and programs.

\section{Data types in FITS} \label{app:fitsdatatypes}
Data and metadata in FITS \textit{may} be represented in many forms depending on the context.

In headers, metadata is defined by ASCII-text representations of literal string, logical or numerical constants.

In images, data is represented by bits which, depending on the value of the \texttt{BITPIX} keyword, \textit{must} be interpreted as characters, integers or floating-point values. Integers and floating-point data \textit{must} be represented in different ways depending on the desired range of values, accuracy and size constraints.

In binary tables, data is also represented by bits. However, different fields \textit{may} have different data types, which are defined through their respective \texttt{TFORM}\textit{n} keywords. Similar to images, integer, floating-point, complex and array data in binary tables \textit{may} be represented in different ways depending on the desired range of values, accuracy and size constraints. Specifically, a field \textit{may} contain data of one of following types:
\begin{itemize}
    \item Logical: interpreted as true or false (\texttt{TFORM} is \texttt{'L'}).
    \item Binary: interpreted as literal bits (\texttt{'X'}).
    \item Character: interpreted as the restricted set of 7-bit ASCII-text characters (\texttt{'A'}).
    \item Integer: interpreted as signed 16-bit (\texttt{'I'}), 32-bit (\texttt{'J'}) or 64-bit (\texttt{'K'}) integers, or unsigned 8-bit integers (\texttt{'B'}).
    \item Floating-point: interpreted as 32-bit single-precision (\texttt{'E'}) or 64-bit double-precision \texttt{'D'} floating point data.
    \item Complex: interpreted as 32-bit single-precision (\texttt{'C'}) or 64-bit double-precision \texttt{'M'} complex data.
    \item Array descriptor: interpreted as two 32-bit (\texttt{'P'}) or 64-bit (\texttt{'Q'}) integers. These two integers are used to essentially define the starting point and length of a variable-length array (VLA), which is then stored in a supplemental data area (heap) following the main data table. The \texttt{TFORM} of a VLA column \textit{must} define two characters, the first for the type of array descriptor, and the second for the data type of the actual array stored in the heap. This means that a VLA \textit{may} be of any of the types defined previously.
\end{itemize}
For fields of types other than array descriptor, their entries are assumed by default to contain a single value. However, those fields \textit{may} define their entries as arrays of values; the number of elements (array length) is specified by a non-negative integer in \texttt{TFORM} which precedes the type character, meaning that all entries in such a field \textit{must} have the exact same number of elements (this limitation is avoided in VLAs, due to the reasons elucidated above). Multidimensional data \textit{may} also be specified by a \texttt{TDIM} keyword, which also applies to all entries in the field; this has some limitations presented in Appendix~\ref{app:multidimensionaldata}.

\section{Multidimensional data in AOT} \label{app:multidimensionaldata}
While the FITS specification \textit{technically} supports storing multidimensional data in binary tables, this feature has many limitations. The specification defines that, in order for a certain field to contain multidimensional entries, this field \textit{must} have a corresponding \texttt{TDIM}\textit{n} keyword which defines the exact dimensions of its arrays. While the array itself is still stored as a one-dimensional set of data, knowing its intended dimensions allows for a proper interpretation of the data.

Since we \textit{must} only define one such keyword per field, a direct consequence is that all the entries in that field \textit{must} have the exact same dimensions\footnote{There is one exception to this: fields containing multidimensional data formatted as variable-length arrays. In this case, their entries \textit{may} either adhere to the specified dimensions or be empty entries (length 0). This is a very obscure detail which is currently unsupported by most FITS tools. Regardless, this feature is not particularly useful for the situation at hand, as VLAs provide no mechanism for defining different dimensions per entry, and thus there is no meaningful space savings to be had.}. Using AOT as an example, this means that if we wanted to store the slopes of a certain wavefront sensor as a multidimensional array inside the binary table, slopes of all further wavefront sensors would also need to respect those exact dimensions. However, it is not guaranteed that they would have the same dimensions, as the number of subapertures might not be the same, or we might be recording at a different frequency. Therefore, it is clear that the standard implementation of multidimensional data in binary tables is not versatile enough to fit AOT's needs.

Since we had to store multidimensional data in some way, we considered three ways of getting around this limitation:
\begin{enumerate}
    \item Use \texttt{TDIM}\textit{n} keywords and, for each field, pick a large enough set of dimensions such that the data from all entries can fit inside those dimensions. Fill the remaining spaces with some sort of filler value that is ignored when reading.
    \item Use variable-length arrays without \texttt{TDIM}\textit{n} keywords, and add an extra column which specifies the dimensions for each entry.
    \item Save the multidimensional data as separate image extensions that are then referenced in the corresponding table entries.
\end{enumerate}

The first option has multiple issues. First off, it can result in large amounts of wasted storage space given that a user could be forced to write a small array as a much larger one, in order to respect the dimensions that were set by another array in that field. On top of that, defining an appropriate filler value could be difficult given how diverse the data can be. It is also completely unintuitive for any users that would see such data without context, as the FITS specification would likely lead them to interpret all entries as having the dimensions specified by \texttt{TDIM}\textit{n}. Finally, the implementation complexity for this would be relatively large, forcing reading and writing tools to filter and reshape possibly huge amounts of data in order to correctly interpret it.

By using VLAs, the second option avoids the wasted space that the first option would create. However, it would force the creation of one extra field for any field that contains multidimensional data, which is a mechanism that is not standardised in FITS. Therefore, the interpretation of this field would be ambiguous for users not familiar with the AOT standard (for example, are the arrays be stored in row-major or column-major order?), which could lead to interchangeability issues.

We ended up choosing the third option. The major deciding factor is that saving multidimensional data in image extensions is a very well-defined mechanism, since it is fully described in the FITS specification. Using image extensions also allows us to leverage their already established features and versatility, such as saving metadata in its headers or using data scaling features. Another advantage is that FITS tools are usually already prepared to deal with image extensions and thus they can easily and unambiguously interpret them without requiring any AOT-specific conventions. Finally, this approach also allows us to save multidimensional data externally through FITS image files, which is a very common mechanism. The biggest downside of using this approach is that it requires AOT to use a non-standard cross-referencing mechanism. However, the format already required a similar mechanism to deal with cross-referencing between binary tables (see Appendix~\ref{sapp:crossreferencing}), therefore this did not represent a significant increase in AOT's complexity.

It is important to note that some fields in AOT contain one-dimensional arrays that \textit{may} have a length larger than 1 (that is, lists). While such data could reasonably be stored as one-dimensional image extensions, we deem doing so would create unnecessary clutter in the file structure. This is because 1) these sets of data tend to be relatively small and fairly simple, so they are unlikely to benefit significantly from the metadata of image extensions and 2) the standard VLAs are able to fully describe these lists without requiring any AOT-specific convention. Therefore, in fields where one-dimensional arrays are expected, AOT uses VLAs.

\section{AOT Conventions} \label{app:conventions}
In this appendix we will define a set of AOT-specific conventions that apply to the entirety of this paper. Where relevant, some standard FITS conventions will also be mentioned, but we will not cover the entirety of the FITS specification; it is implicit that all FITS features are supported, unless explicitly defined otherwise.

\subsection{Mandatory FITS Keywords} \label{sapp:mandatoryfits}
The FITS specification defines a set of mandatory keywords \citep[Section 4.4.1.]{chiappetti_definition_2018} for both the primary and extension headers. For simplicity, in the following sections we will only explicitly define AOT-specific keywords for each HDU, but the mandatory FITS keywords are implicitly assumed to be present and appropriately adhered to.

\subsection{Data types}
The FITS specification defines a wide range of possible data types for which we provide an overview in Appendix~\ref{app:fitsdatatypes}. Notwithstanding, in AOT only 4 different types of data are used:
\begin{itemize}
    \item Integer, represented as ``\textit{int}''.
    \item Floating-point, represented as ``\textit{flt}''.
    \item String, represented as ``\textit{str}''.
    \item List of floating-point data, represented as ``\textit{lst}'';
\end{itemize}
Header values are either \textit{str} or \textit{flt}. Image data \textit{may} be \textit{int} or \textit{flt}. Fields in binary tables \textit{may} be \textit{int}, \textit{flt}, \textit{str} (represented as a fixed-length array of characters) or \textit{lst} (represented as a variable-length array of floats). AOT does not specify the precision for any of the data types; this means that a field or image specified as, e.g., \textit{flt} \textit{may} be represented with single (32-bit) or double (64-bit) precision, according to the user's needs (so long as the FITS specification is properly adhered to).

Although the FITS specification technically allows binary table fields of any data type to contain multidimensional arrays, this functionality is not used in AOT and thus all data within AOT binary tables \textit{must} be uni-dimensional (for reasoning, see Appendix~\ref{app:multidimensionaldata}). As a result, all multidimensional data in AOT is contained in image extensions that are cross-referenced via table entries (this mechanism will be explained in Appendix~\ref{sapp:crossreferencing}).

\subsection{Cross-referencing} \label{sapp:crossreferencing}
\begin{figure*}[htbp]
   \begin{center}
   \includegraphics[width=\textwidth]{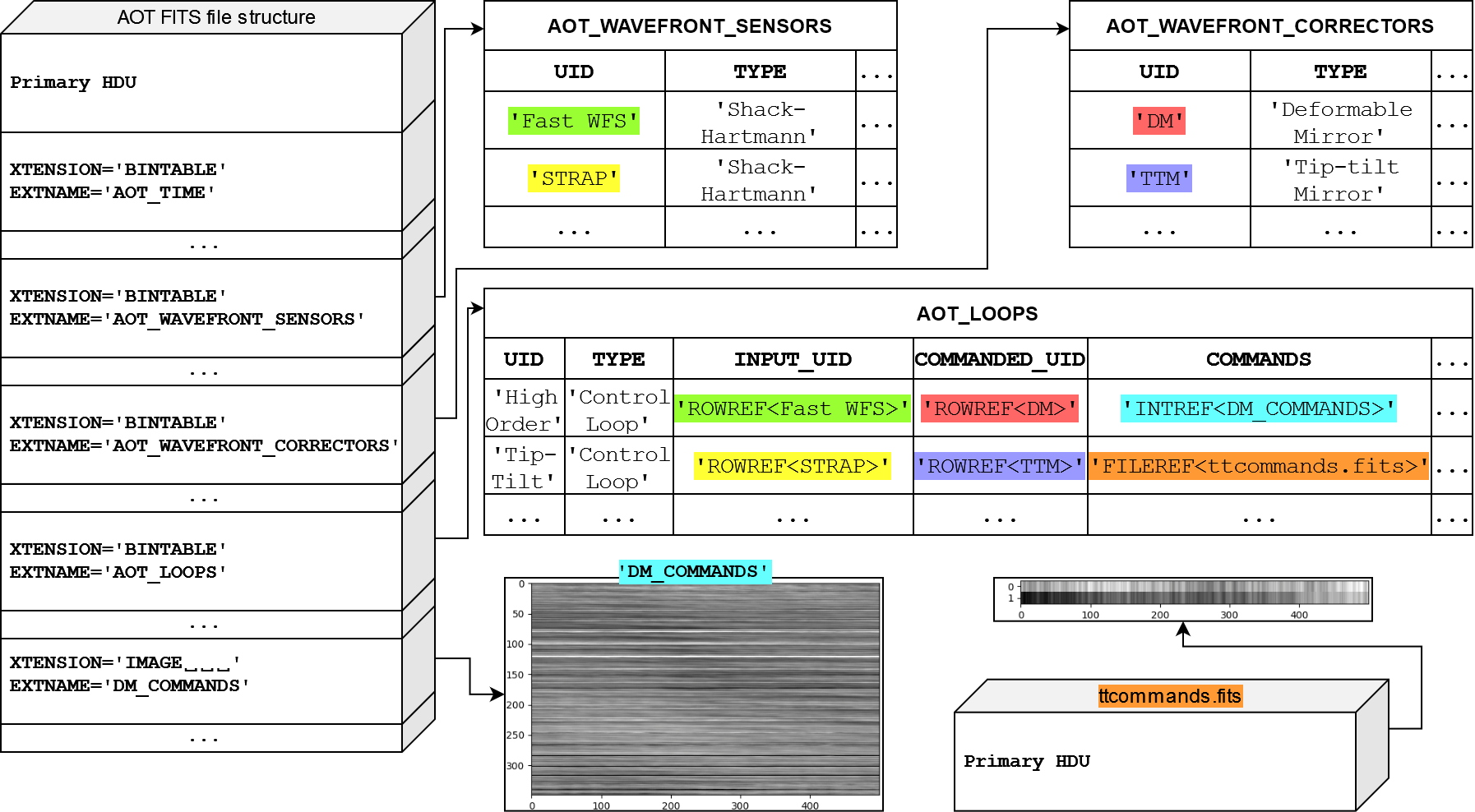}
   \end{center}
   \caption{\label{fig:referencing} 
    A diagram demonstrating how the cross-referencing mechanism in AOT works. In this example, each row in the AOT\_LOOPS table references one row in the AOT\_WAVEFRONT\_SENSORS table, one row in the AOT\_WAVEFRONT\_CORRECTORS table and one image extension. Rows are identified by the string value of their \texttt{UID} fields, while images are identified by the string value of their \texttt{EXTNAME} keyword.}
\end{figure*}

The FITS specification does not provide a standard way of cross-referencing image extensions or table rows. Given that such a mechanism would be convenient for defining relationships between data and avoiding data duplication, AOT defines its own cross-referencing mechanism. A diagram demonstrating its use can be seen in Fig.~\ref{fig:referencing}.

In general, there are two different types of references in AOT: \textbf{row-references} and \textbf{image-references}. Both types of references are done through a single \textit{str} value, in a field that is specified as being a reference. This \textit{str} value is able to uniquely and unambiguously identify the referenced row or image extension. A null \textit{str} in a reference field is interpreted as no reference being made (see Appendix~\ref{sapp:null}).

A row-reference identifies a specific row of a certain binary table. All AOT tables contain one \texttt{UID} field, for which all the entries \textit{must} have a non-null value that is unique within each table (see Sec.~\ref{ss:bintables}). This means that we can uniquely identify any row as long as we know its \texttt{UID} and the table it belongs to. Given that AOT specifies which table is being referenced in every row-reference field, a row-reference is made unambiguously with a \textit{str} value formatted as \texttt{'ROWREF<\textit{row\_UID}>'}, where \texttt{\textit{row\_UID}} is the \texttt{UID} of the row being referenced.

An image-reference identifies a certain image extension. The referenced extension \textit{may} either be present in the AOT file itself (internal image-reference) or in another FITS file (external image-reference), but \textit{must} always follow the specifications of one of the image categories defined in Sec.~\ref{ss:images}. All image extensions present in the AOT file \textit{must} have a unique non-null name (\texttt{EXTNAME} keyword), allowing them to be referenced internally by a \textit{str} value formatted as \texttt{'INTREF<\textit{image\_EXTNAME}>'}, where \texttt{\textit{image\_EXTNAME}} is the name of the image extension being referenced. 

An external image-reference identifies the FITS file where a certain image extension is present, either through a filename (file image-reference) or through a web address (URL image-reference). A file image-reference is made through a \textit{str} value formatted as \texttt{'FILEREF<\textit{filename}.fits>\textit{index}'}, where \texttt{\textit{filename}} is the name of the FITS file where the image data is located. To ensure wide compatibility across most commonly used filesystems, \texttt{\textit{filename}} \textit{must} be case-sensitive and only contain alphanumeric characters, full stops (\texttt{.}), underscores (\texttt{\_}) and hyphens (\texttt{-}). It is \textit{recommended} that, when file image-references are used, the referenced file is somehow made available along with the AOT file, such that it is unambiguously be referenced solely by its name. An URL image-reference is made through a \textit{str} value formatted as \texttt{'URLREF<\textit{url}>\textit{index}'}, where \texttt{\textit{url}} is an URL \citep[as defined in][]{rfc3986} to a FITS file. It is highly \textit{recommended} that this file is publicly accessible and that it is not ephemeral; in other words, the file \textit{should} remain accessible through this URL by anyone for as long as the data is relevant. For both types of external image-references, \texttt{\textit{index}} is a non-negative integer which indicates the position of the HDU that contains the referenced image data (where index 0 is the primary HDU, index 1 is the first extension, and so on). \texttt{\textit{index}} \textit{may} be omitted as long as the referenced image data is present in the first HDU that contains non-null image data. Given the challenges associated with guaranteeing that an external image extension always remains unambiguously accessible throughout the lifetime of some data, we highly recommend to only use external image-references in instances where avoiding data duplication is a major concern.

While the values of both the \texttt{EXTNAME} keywords and the \texttt{UID} entries are freely defined by the user, it is highly \textit{recommended} that these values are human-readable and that they intuitively describe the set of data that they correspond to. This is beneficial to the readability of the file, as it helps making it self-explanatory. For example, we recommend the \texttt{UID} of a Telescope row to be the name of the telescope itself.

\subsection{Null values} \label{sapp:null}
Since AOT defines a wide set of data, we expect that users might not want to fill all of the available fields. Therefore, with the exception of the mandatory FITS keywords (Appendix~\ref{sapp:mandatoryfits}), the fields necessary for uniquely identifying references (Appendix~\ref{sapp:crossreferencing}) and other keywords defined as strictly mandatory in Secs.~\ref{ss:primary} or \ref{ss:bintables}, all values are considered to be nullable.

While the FITS specification allows for keywords in headers to have values of indeterminate data type (undefined keywords), to avoid confusion this is strictly disallowed in AOT. Therefore, null strings \textit{must} be represented as zero-length arrays of characters (\texttt{KEYWORD =\textquotesingle\textquotesingle}, see \citeauthor[Sec.~4.2.1.1]{chiappetti_definition_2018} for a detailed distinction).

FITS specifies that null values in binary tables are represented in different ways depending on their specific data type:
\begin{itemize}
    \item Null \textit{str} data \textit{must} be represented by an array of characters where the first character is the ASCII NULL character (0x00), in other words, a zero-length null-terminated string.
    \item In order for null \textit{int} data to exist in a certain field, that field \textit{must} have a \texttt{TNULL}\textit{n} value specified by the user; all entries containing that exact value are then interpreted as null. Given that all integer fields in the AOT standard describe counts, negative integer values inherently have no meaning and thus are \textit{recommended} for this purpose (assuming a signed representation).
    \item For null \textit{flt} values, the IEEE special value NaN \textit{must} be used.
    \item Null \textit{lst} are specified by having an array length of 0, and \textit{should} be interpreted as an empty array rather than a non-existing one (see \citeauthor[Sec.~7.3.5. and 7.3.6.]{chiappetti_definition_2018}). 
\end{itemize}

If a user does not intend to provide data for a field that references an image, it is preferable to use a null reference, rather than referencing an image extension that contains no data.

\subsection{Date and time representation} \label{sapp:time}
Date and time data is vital for interpreting AO telemetry data; accordingly AOT supports associating date and time data to all time-sensitive and/or time-varying sets of data. All date and time data in AOT \textit{must} be in the UTC time standard, and \textit{must} be represented in one of two ways (depending on the specific field): 1) Based on the ISO 8601 format or 2) Unix timestamp.

ISO 8601 provides a set of unambiguous ways of representing calendar dates and times in a human-readable format. In AOT, we use a ISO 8601 derived format where date and time are represented as a string in the pattern <date>T<time> with <date> being specified as YYYY-MM-DD and <time> specified as hh:mm:ss[.SSSSSS] (the decimal part, in square brackets, is \textit{optional}). This format is used where a single timestamp is expected, for its human-readability. No other ISO 8601 formats are supported.

In situations where a set of timestamps is expected (for example, timestamps related to time-varying data) the Unix time format (also known as POSIX or Epoch time) is used instead. This format represents date and time as a floating-point value containing the number of seconds that have elapsed since 00:00:00 UTC on 1 January 1970, with decimal digits being allowed. Dates past 03:14:08 UTC on Tuesday, 19 January 2038, cannot be represented by 32-bit floats (due to overflow); in such cases, 64-bit floats \textit{must} be used.

\subsection{Geometric transformations} \label{sapp:geometry}
\begin{figure}
    \centering
    \includegraphics[width=\columnwidth]{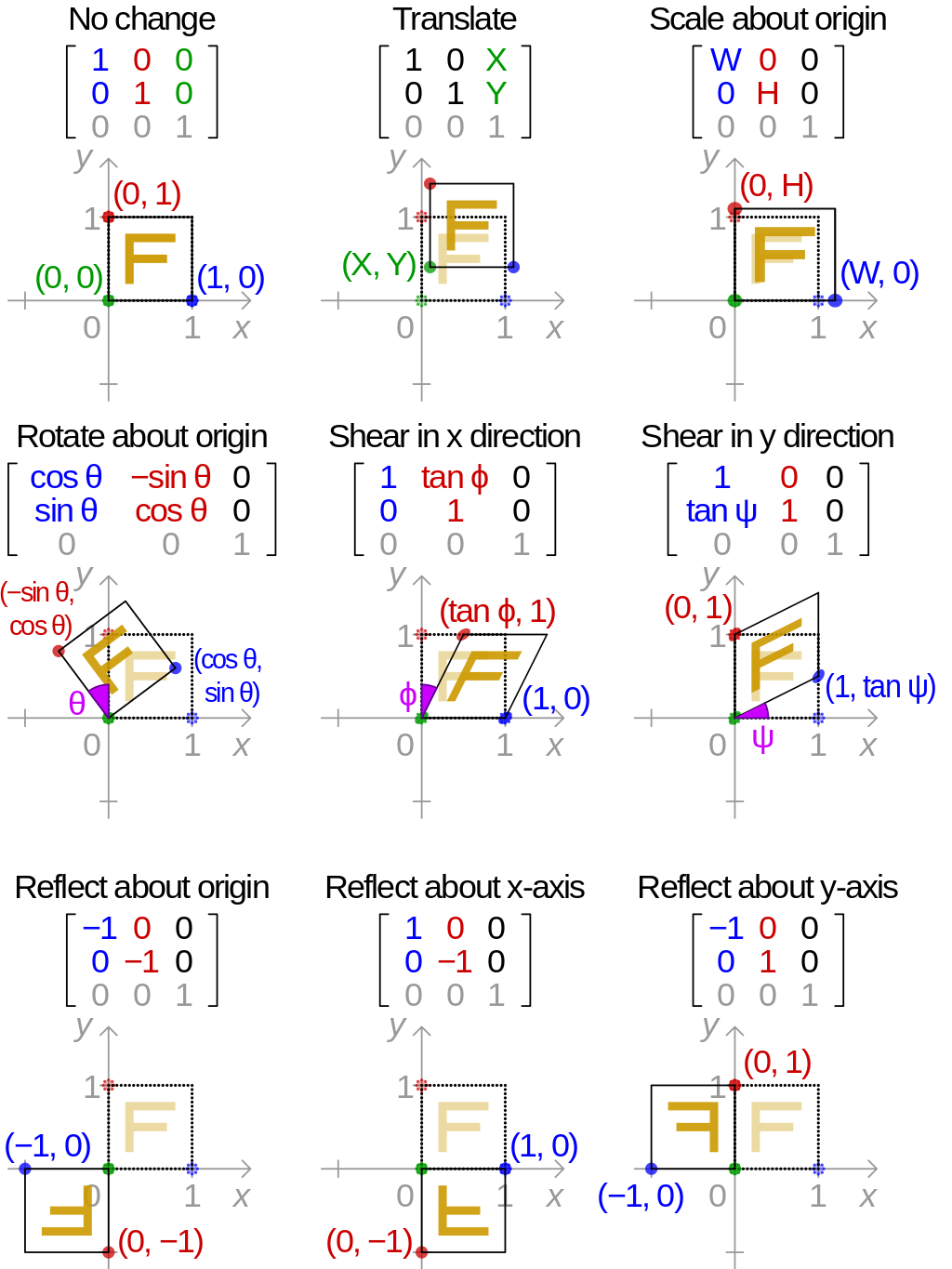}
    \caption{\label{fig:transformations}
    Diagram describing how different affine transformations can be performed via 3-by-3 matrices. Illustration by Cmglee, distributed under the Creative Commons Attribution-Share Alike 3.0 Unported license \citep{wiki:geometric}.}
    
\end{figure}

Any combination of translation, reflection, scale, rotation and shearing (affine transformations) can be described via a single \(3 \times 3\) matrix \(M\) such that \(P' = MP\), where \(P\) is a \(\begin{bmatrix}x & y & 1 \end{bmatrix}\) vector (with \(x\) and \(y\) being the original horizontal and vertical coordinates, respectively) and \(P'\) is a \(\begin{bmatrix}x' & y' & 1 \end{bmatrix}\) vector, where \(x'\) and \(y'\) are the transformed coordinates. A diagram exemplifying this mechanism can be see in Fig.~\ref{fig:transformations}.

All geometric information in AOT \textit{must} be described relative to the same reference origin point (defined by the user), from which transformations \textit{may} occur. For example, if the geometry of the system is defined with the telescope pupil as the reference point, the \texttt{TRANFORMATION\_MATRIX} on the corresponding AOT\_TELESCOPES row will be a 3-by-3 identity matrix. Then, a wavefront sensor may define a 90\degree\ rotation from the pupil with a \(\begin{bsmallmatrix}
  0 & -1 & 0\\
  1 & 0 & 0 \\
  0 & 0 & 1 \\
\end{bsmallmatrix}\) matrix. The detector used by that wavefront sensor, if not rotated in regards to it, would define its geometry by using the exact same matrix.

\begin{acknowledgements}
This project has received funding from the European Union’s Horizon 2020 research and innovation programme under grant agreement No. 101004719 (OPTICON–RadioNet Pilot) and from the FCT - Fundação para a Ciência e a Tecnologia through the grant CENTRA UIDB/00099/2020.

C. M. Correia acknowledges the financial support provided by FTC with grant 2022.01293.CEECIND.
\end{acknowledgements}
\bibliography{aots.bib}

\end{document}